\documentclass[preprint2]{proto}
\usepackage{times}
\usepackage{xcolor}
\usepackage{graphicx}
\usepackage{float}
\usepackage{dblfloatfix}
\usepackage{amsmath}
\usepackage{amsfonts}
\usepackage{hyperref}
\hypersetup{
    colorlinks=true,
    linkcolor=red,
    filecolor=magenta,     
    citecolor=blue,
    urlcolor=cyan,
    }

\usepackage[switch]{lineno}

\graphicspath{{./}{Figures/}}

\begin{document}

\title{\textbf{\LARGE Neutral gas coma dynamics: modeling of flows and attempts to link inner coma structures to properties of the nucleus}}

\author {\textbf{\large Raphael Marschall}}
\affil{\small\em Southwest Research Institute, USA\\ Observatoire de la Côte d'Azur, France}

\author {\textbf{\large Björn J. R. Davidsson}}
\affil{\small\em Jet Propulsion Laboratory / California Institute of Technology, USA}

\author {\textbf{\large Martin Rubin}}
\affil{\small\em University of Bern, Switzerland}

\author {\textbf{\large Valeriy Tenishev}}
\affil{\small\em University of Michigan, USA}

\begin{abstract}

\begin{list}{ } 
{\rightmargin 0in}
{\leftmargin -1in}
\baselineskip = 11pt
\parindent=1pc
{\small 
{\small\em \centerline{Accepted: January 23, 2023} }
\vspace{0.25cm}

Deriving properties of cometary nuclei from coma data is of significant importance for our understanding of cometary activity and has implications beyond. 
Ground-based data represent the bulk of measurements available for comets. Yet, to date these observations only access a comet’s gas and dust coma at rather large distances from the surface and do not directly observe its surface or even the outgassing layer. 
In contrast, spacecraft fly-by and rendezvous missions are one of the only tools that gain direct access to surface measurements. 
However, these missions are limited to roughly one per decade. 
We can overcome these challenges by recognising that the coma contains information about the nucleus's properties. 
In particular, the near-surface gas environment is most representative of the nucleus. 
It can inform us about the composition, regionality of activity, and sources of coma features and how they link to the topography, morphology, or other surface properties. 
The inner coma data is a particularly good proxy because it has not yet, or only marginally, been contaminated by coma chemistry or secondary gas sources (e.g., from icy grains released into the coma), and can retain fine structure which need time to dissipate.
Additionally,  when possible, the simultaneous observation of the innermost coma with the surface provides the potential to make a direct link between coma measurements and the nucleus. 
If we hope to link outer coma measurements obtained by Earth-based telescopes to the surface, we must first understand how the inner coma measurements are linked to the surface. 
Numerical models that describe the flow from the surface into the immediate surroundings are needed to make this connection. 
This chapter focuses on the advances made to understand the flow of the neutral gas coma from the surface to distances up to a few tens of nuclei radii. 
The current state of research on linking the inner gas coma properties and structures to the nucleus is explored, describing both simple/heuristic models and state-of-the-art physically consistent models. 
The model limitations and what they each are best suited for is discussed. 
In the end, the different approaches are compared to spacecraft data, and the remaining knowledge gaps and how best to address them in the future are presented.\\\\\\

}
\end{list}
\end{abstract}  

\section{\textbf{Introduction}}\label{sec:introduction}
Comets are thought to be icy leftovers from planet formation, either planetesimal themselves or direct descendants of the former.
For that reason, they are widely considered to have retained information about the early Solar System and can inform our understanding of planet formation.
While their interiors have likely retained their primordial properties, the same cannot be said for their surfaces \citep[e.g.,][]{Jutzi2020Icar}.
Cometary surfaces can be considered heavily evolved by numerous processes such as e.g., irradiation, impacts, thermal processing, and sublimation-driven activity.
For more on the structure and properties of the surface, see Chapter XX in this volume.

Because the pristine interior of comets is not easily accessible directly, we turn our gaze to the gas and dust comae, which can be studied with spacecraft and ground-based telescopes. 
The Deep Impact mission \citep{AHearn2005SSRv} stands out for probing the subsurface of 9P/Tempel 1 with an impactor and visiting the first hyperactive comet (103P/Hartley 2).
But this bridge, from the interior/surface to the comae, requires us to devise methods to link the coma properties to the surface/interior.
We need to understand the dynamics of the gas and dust from the surface to a spacecraft or the distances observed with ground-based telescopes.
This chapter will describe the current state of the art in modelling the gas dynamics within the first few nucleus radii above the surface (corresponding to a few tens of kilometres in the case of a nucleus with a typical radius of a few kilometres) and critical insights from the past decade of research and spacecraft missions.

To date, only six comets (1P/Halley, 19P/Borrelly, 9P/Tempel~1, 67P/Churyumov-Gerasimeko, 81P/Wild 2, and 103P/Hartley 2) have been visited by spacecraft, which resolved their nuclei.
Spacecraft observations provide a detailed, high spatial and temporal resolution of the surface and surrounding coma but are limited to a few target comets ($\sim\!1-2$ per decade).
ESA's Rosetta mission \citep{Glassmeier2007} has given us the most recent and detailed picture of cometary evolution by following comet 67P/Churyumov-Gerasimenko (hereafter 67P) through its perihelion for over two years. 
In fact, apart from Rosetta, all previous comet missions have been fly-bys and thus did not cover the baseline to study temporal changes.
Only Rosetta provided a long time data set to study the temporal variability and evolution of the coma in detail.
Previous missions were able to observe comets for several rotation periods of the nucleus and were able to find periodicity in their activity.
For example, the combination of Stardust NExT’s exploration of 9P/Tempel~1 one full apparition after the Deep Impact experiment showed significant changes  on the surface \citep{Veverka2013Icar}.

Comets are more easily and frequently observed using ground- and space-based telescopes.
In contrast to spacecraft, the spatial resolution is much lower, but we can observe many more comets and thus sample their diversity.
Though the nuclei are not resolved in telescopic observations, the comae and tails are.

Both spacecraft and ground-/space-based telescopes thus provide complementary data sets that require consolidation.
In this chapter, we focus on the near nucleus coma (within the first few nucleus radii of the surface).
The three main reasons that motivate the study of this region are:
\begin{enumerate}
    \item We may link spacecraft measurements from the coma to the surface.
    \vspace{-0.2cm}
    \item Understanding the innermost coma is a prerequisite to understanding ground-based observations and linking those measurements to the nucleus. I.e. we first need to understand the near nucleus coma to interpret ground-based data.
    \vspace{-0.2cm}
    \item The gained knowledge of this region allows us to make predictions for future comet missions and assess hazards for spacecraft operating in that region.
\end{enumerate}

Though we will touch on the issue of dust in the gas flow, dust dynamics is not the main focus of this chapter.
Instead, we refer the reader to Agarwal et al. in this volume and \cite{Marschall2020c} for detailed reviews of the state of the art in dust coma research.
We will however discuss how dust can alter the properties of the gas flow but will leave the rest to the two references above.

Spacecraft- and Earth-based telescopes measure gas column densities along the line of sight.
This is done indirectly through the measurement of emission lines of different gas species \citep[e.g.,][]{Feaga2007Icar,Biver2019A&A} or absorption of starlight during occultations when in orbit with the comet \citep[e.g.,][]{Keeney2019AJ}.
A spacecraft, when embedded in a coma, can additionally measure the local gas densities \citep[e.g.,][]{Hassig2015Sci}.
Both quantities, local gas densities, and line-of-sight column densities can be used to derive the parameters of the gas flux at the surface.
This includes the gas production rate globally and the distribution of sources at the surface.
Further, the relative abundances in the coma bear information on the composition of the ices in the nucleus \citep[e.g.,][]{Marboeuf2014Icar,Prialnik1992ApJ,Herny2021}. 

In this chapter we will focus on two crucial questions of linking inner coma measurements to the surface:
\begin{enumerate}
    \item How can we confidently derive the gas production rate of different species and thus the volatile mass loss from coma measurements?
    \vspace{-0.2cm}
    \item Can we determine if coma structures (inhomogeneities in density, often referred to as ``jets'') are reflective of a heterogeneous nucleus, or are mere emergent phenomena in the gas flow due to, e.g., the complex shape of the nucleus?
\end{enumerate}

We will only focus on the inner coma/near environment for this review.
There is no strict definition of the inner coma, but here we consider it the region within which the major gas species (H$_2$O, CO$_2$, and CO) accelerate and do not yet experience any substantial loss through chemical reactions (ionization, ion-neutral reactions, etc.).
These chemical processes act on tens of thousands of kilometres and will make a notable dent in the neutral gas profile \citep[e.g.,][]{Shou2016ApJ}.
The typical extent of the acceleration region is of the order of ten nucleus radii \citep[a few 10s of kilometres for a typical comet;][]{Tenishev2008,Shou2016ApJ,Zakharov2018b}.
This region is typically only accessible with spacecraft missions and not by ground-based observations.

We ultimately want to understand how measurements at larger distances to the nucleus obtained with ground- or space-based telescopes can be linked to the nucleus.
But, before we can understand the link between those measurements and the nucleus, we first need to understand how the near environment can be linked to the surface.
Therefore, we dedicate this chapter to the advances of the latter.

We also focus here on the inner coma because of the recent wealth of spacecraft data - from Rosetta and Deep Impact.
The close distances to the source region of the gas also provide the biggest chance to link the coma to the surface unambiguously.

The second question posed above is controversial as it has been known for some time that it is theoretically possible to produce structures in the coma from a homogeneous but non-spherical nucleus.
Moreover, inhomogeneous spherical and homogeneous aspherical nuclei may lead to the similar structures in the gas coma \citep[e.g.,][]{Zakharov2008}.
The chapter by \cite{Crifo2004} in Comets II left us at that crossroad.
At the time, the only modelling including an actual comet shape and data comparison had been done for 1P/Halley.
Since then, we have added five more comets (19P, 81P, 9P, 103P, and 67P) to help us understand the gas flow from cometary nuclei.
At the time of the previous book, the modelling of the inner coma was still primarily theoretical.
There was a large amount of work done which explored active spots on or inhomogeneous outgassing from spherical nuclei \citep[e.g][]{KomleIp1987, Kitamura1990, Knollenberg1994, Crifo1995, CrifoRodionov1997} as well as homogeneous nuclei with complex shapes such as ellipsoids and beans \citep[e.g.,][]{Crifo1997a, Crifo1999, CrifoRodionov2000, Crifo2002bAAF}.
\cite{Crifo2004} had to leave the question as to what drives inner coma structures open, and thus we intend to revisit this question in this chapter and provide some answers.

We will show in Section~\ref{sec:heuristicModels} that the gas production rates can be reasonably safely estimated using heuristic models, at least for some comets.
In Section~\ref{sec:stateOfTheArtModels} we will present the state of the art of physical gas coma models and argue in Section~\ref{sec:emergentStructures} that the evidence point to the fact that observed coma structures do not require a heterogeneous nucleus.
Rather redistribution of material across the nucleus surface is sufficient to explain most regional heterogeneity of the observed activity.
Overlain on these regional levels of activity is topography and the irregular shape of the nucleus that affect the flows through focussing and de-focussing.
We will conclude this chapter by giving an outlook and discussing open questions (Section~\ref{sec:outlook}).

\section{Coma structures definitions} \label{sec:structures}

\begin{figure*}
	\includegraphics[width=\textwidth]{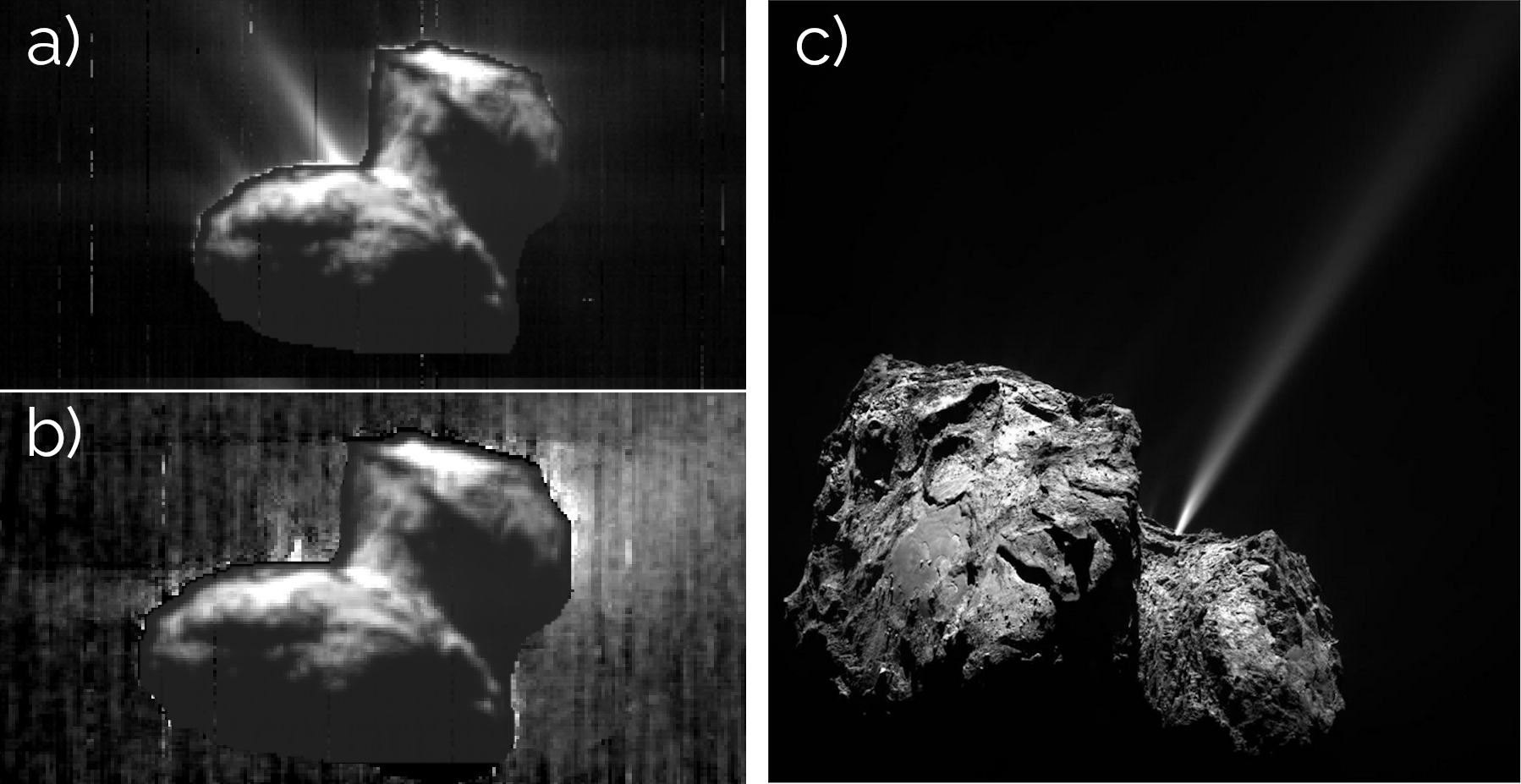}
	\caption{Panels a) and b) show composite images of the comet 67P/Churyumov-Gerasimenko's nucleus at 1.095 $\mu$m, superimposed with the water column density (a) and CO$_2$ column density (b) observed by Rosetta/VIRTIS-M \citep[both panels adapted from][]{Migliorini2016}\protect\footnotemark. Panel c) shows comet nucleus observed by Rosetta/OSIRIS \citep[panel presented in][]{Vincent2016MNRAS}\protect\footnotemark. It shows the nucleus and the scattered light of a collimated feature, a ``jet'' of dust grains.}
	\label{fig:comaStructures} 
\end{figure*}
\addtocounter{footnote}{-1}
\footnotetext[\thefootnote]{Credit: Courtesy of Alessandra Migliorini through private communication. The figure was adapted from Migliorini et al., A\&A, 589, A45, 2016, reproduced with permission © ESO.}
\addtocounter{footnote}{1}
\footnotetext[\thefootnote]{NAC image taken on 2015-07-29 13:25:28 UTC; Credit: ESA/{\allowbreak}Rosetta/{\allowbreak}MPS for OSIRIS Team MPS/{\allowbreak}UPD/{\allowbreak}LAM/{\allowbreak}IAA/{\allowbreak}SSO/{\allowbreak}INTA/{\allowbreak}UPM/{\allowbreak}DASP/{\allowbreak}IDA published under the Creative Commons license CC BY-SA 4.0.}

The main property to differentiate structures in the gas flow is the spatial scale. 
There are large/global scale (larger than the nucleus) structures and fine structures (much smaller than the scale of the nucleus). 

A good example of the former large-scale gas flow structure is the CO$_2$ column density distribution in the coma of comet 103P/Hartley 2 as observed during the Deep Impact eXtended Investigation \citep[DIXI,][]{Protopapa2014} shown in Fig.~\ref{fig:comaStructures}.
This feature does not appear to have a confined source region but rather covers a significant fraction of the smaller lobe of comet 103P.
These larger-scale structures reflect the global parameters, such as the total gas production rate, asymmetry and composition in gas production, and the large-scale geometry of the nucleus (amongst others its shape and rotation state).

In contrast, the fine structures reflect very local features of the nucleus including its topography.
A good example, though only indirectly observed through the reflectance of dust particles in the gas flow, are highly collimated features observed at comet 67P/Churyumov-Gerasimenko during the Rosetta mission \citep{Vincent2016MNRAS}.
These events can be associated with outbursts.
In such cases, we might refer to these features as a ``jets''.
The use of the word ``jet'' is controversial though, mainly because it has a strict physical interpretation but is often used very liberally to describe any collimated feature in the coma \citep[see, e.g.,][]{Vincent2019}.
Therefore, in some parts of the literature, any inhomogeneity in the coma that appears to be collimated will be identified as a ``jet''.
We argue that a ``jet'' should have a narrower definition, which is closer to a physical understanding of the word.
At least the following two properties should be satisfied.
A ``jet'' represents a gas stream with i) a clear boundary with respect to ambient flow and ii) outflowing from a source much smaller than the size of the nucleus (Fig.~\ref{fig:comaStructures}a/c).
If the source region covers, e.g., an entire hemisphere it would not be ``jet'' even though the resulting feature might appear bounded. 
In this example, we would suggest the less implicating term ``stream'' (The CO$_2$-feature in the panel b) of Fig.~\ref{fig:comaStructures} nicely fits that). 
A counter-example to a bounded ``jet'' is the expansion into a solid angle of $2\pi$ which would rather be regarded as a ``plume''. 
We will discuss in Section~\ref{sec:emergentStructures} why this nomenclature can be extremely misleading and that except for outbursts most features in the coma don't warrant the label ``jet''. 
We should also note that for historical reasons the word ``jet'' is often used to describe features observed in the outer coma of ground-based data and is used descriptively.
Here, we specifically encourage a more specific use for inner coma structures because there we have at least the possibility to more accurately distinguish between these terms.

With increasing distance from the surface, the flow expands and the local density decreases.
The mean free path (MFP) of the molecules, therefore, becomes large and therefore the flow gradients become smoothed.
At large distances to the nucleus the increased rarefaction causes the fine structures of the flow to vanish. 
This is why coma structures on large distances will only reflect the global characteristics of the neutral gas flow. 
Nevertheless, even these global characteristics are of interest since they allow capturing the general properties of gas emission (e.g., total gas production rate) and therefore those of the nucleus.


\section{Heuristic models}\label{sec:heuristicModels}
A heuristic model by its nature sacrifices physical rigor for simplicity, and therefore computation speed.
It attempts to simplify a problem by neglecting complexities deemed unnecessary to derive certain properties.
If physical models (Section~\ref{sec:stateOfTheArtModels}) were computationally cheap there would not be any justification at all to turn to heuristic models.
They can be useful to obtain rough estimates and thus provide a "sanity check" on physical models.
Some inconsistencies or uncertainties of heuristic models are easy to spot, other limitations are not immediately obvious.
It is therefore important to understand the limitations of heuristic models and not apply them to inappropriate situations or the resulting interpretation of the respective data is likely to be wrong.
As we will see some heuristic models can be useful but they also immediately show the need for physically accurate models which we will discuss in Section~\ref{sec:stateOfTheArtModels}.

The most famous heuristic model in cometary comae research is the so-called ``Haser model''  \citep{Haser1957}. 
It assumes free molecular (i.e. collision-less) radial flow and is based on the conservation of the number of particles.
It also takes into account chemical processes like photo-dissociation.
In its simplified form, where chemical processes can be neglected, the model can be greatly simplified.
In this case it links the total gas production rate, $Q$, to the local gas density $n$ via
\begin{equation} \label{eq:Haser}
    Q = 4 \pi r^2 n v \quad,
\end{equation}
where $r$ is the distance to the nucleus, and $v$ is a constant gas speed.
The equation above also assumes isotropic expansion.
Non-isotropic flows will also reach radial expansion at a constant speed and at that point the above expression is valid in a directional sense, conserving the directional mass flow.
In its full form, the model includes the photo-chemical destruction of parent molecules and can be rewritten to track daughter species \citep{Combi2004}.
For a comet at 1~au these chemical processes act on tens of thousands of kilometres and will make a notable dent in the neutral gas profiles only on those scales \citep[e.g.,][]{Shou2016ApJ}.
These processes do not dominate on the short timescales of the near nucleus environment.
The flow can also be confined to, e.g., a half sphere (for instance the sunward-side) by modifying the solid angle from $4\pi$ to $2\pi$ steradian.
In this sense, this is the simplest gas model one might think of.
Although it is still the most commonly used approximation, owing to its apparent simplicity, it is physically adequate only at distances when the flow is expanding radially at a constant velocity.
To be precise the flow generally expands radially and reaches 90\% of terminal velocity at around ten nucleus radii \citep{Zakharov2018b,Gerig2018Icar}.
This model cannot capture the dynamics close to any nucleus while the gas accelerates and simultaneously cools due to the associated expansion.
At these short distances to the surface ($< 10~R_N$), effects from the nucleus shape also still play an important role.
It should therefore not be used in the immediate vicinity of the nucleus.
But, Eq.~\ref{eq:Haser} can be rather safely used between distances of ten nucleus radii and ~ten thousand kilometres.
Beyond the latter chemical reactions need to be accounted for.

For early data from the Rosetta mission, this model seemed to provide reasonable estimates of the gas production rate using the relationship in Eq.~\ref{eq:Haser} and variations in the solar zenith angle \citep{Bieler2015EPS}.
At that point in the mission, 67P was still beyond 3~au from the Sun, and Rosetta at $\ge10$~km cometocentric distance thus outside the gas acceleration region \citep{Tenishev2008,Zakharov2018b,Zakharov2023}.
We will come back to the question, of whether the ``Haser model'' or other heuristic models are useful to at least estimate the gas production rate.

Recently two more models have appeared combining physics-based and heuristic approaches. 
In the first model, \cite{Fougere2016a} used the local gas densities from the Rosetta's `Rosetta Orbiter Spectrometer for Ion and Neutral Analysis' \citep[ROSINA,][]{Balsiger2007SSRv} instrument to perform a spherical harmonics fit of the data and constrain the surface gas emission distribution.
Regions that overlap due to the concavities of the shape of 67P are ignored.
These surface distributions in conjunction with local illumination were the initial conditions for a 3D kinetic modelling of the dusty gas coma. 
The modelling was done using the Adaptive Mesh Particle Simulator (AMPS), which is a general-purpose Direct Simulation Monte Carlo (DSMC) model \citep{Tenishev-2021-JGR}.
We refer interested readers to \cite{Bird1994} and \cite{Bird2013} for more detail on the DSMC method.

This validation step with a physical model is important because it gives some confidence in the result.
We cannot be confident that the solution is unique because the surface-emission distribution is prescribed a ``spherical'' form.
This approach will show large-scale (e.g., north-south) distribution of gas sources but does not seem adequate to link the surface activity to morphological differences on the surface.

This ``spherical harmonics model'' has not only been used to estimate the surface-emission distribution but also the global gas production rate of the major species along the orbit of 67P \citep{Fougere2016b,Combi2020}.
While the surface-emission distribution contains considerable ambiguity, the global gas production rate overlaps with the results from other approaches \citep[e.g.,][]{Laeuter2020,Marschall2020b}.

The second model \citep{Kramer2017,Laeuter2019,Laeuter2020} assumes that each surface facet of a shape model (here 67P) is an independent gas source.
The gas outflow from each facet is described with an opening angle and then follows essentially collision-less outflow according to \cite{Narasimha1962}.
The different sources from neighbouring facets do not interact and simply contribute linearly to the gas densities at the spacecraft.
In this sense, it is a ``Haser'' type model that also takes into account the shape and we shall refer to it as ``Haser+shape model'' (to some extent similar to \citet{Bieler2015EPS}).
As with any heuristic model, it results in some nonphysical results, e.g., extremely high gas speeds \citep{Kramer2017}.
The model included an assumption on the coma temperature (200 K, 100 K, and 50 K), whereas \cite{Tenishev2008} obtained much lower temperatures in the 30 to 10K range at distances between 10 km and 100 km, which may in part be responsible for this discrepancy. 
Later models \citep{Laeuter2019} then used the modelled velocities by \cite{Hansen2016MNRAS} as input.

Gas flows from different sources cannot be assumed to be independent.
Though this can be true in very rarefied cases, it is rarely true, even for a comet with comparably weak activity as 67P.
It has long been known that gas sources close to each other, e.g., two jets, interact with each other and the local gas density in the coma is not simply a linear combination of the gas densities of the two isolated jets \citep[e.g.,][]{DankertKoppenwallner1984}.
Further, the ``Haser+shape model'' cannot reproduce surface-emission maps from physical coma models \citep[see appendix A of ][]{Marschall2020a}.
Though this makes it unclear if the activity maps from this model are reliable, the global gas production rates using this model \citep{Laeuter2019,Laeuter2020} should be fairly good estimates.
Importantly, \cite{Marschall2020a} point out that there is a physical resolution limit to detecting heterogeneous emission distributions.
This resolution limit to detecting heterogeneous emission distributions stems from the flow viscosity (which is connected with the MFP of the molecules) close to the surface.
Viscous dissipation blurs fine structures of the flow and therefore the underlying information of the boundary conditions which caused these structures.
For a comet like 67P, this resolution limit lies at several hundred meters.
This resolution limit also prevents physical models from determining the activity map to arbitrary accuracy.
Finally, by not taking illumination into account, the ``Haser+shape model'' is also much less suited to reproduce `Microwave Instrument for Rosetta Orbiter' \citep[MIRO,][]{Gulkis2007} and `Visual IR Thermal Imaging Spectrometer' \citep[VIRTIS,][]{Coradini2007SSRv} line-of-sight observations compared to ROSINA data, which were obtained predominantly in a terminator orbit.

\cite{Marschall2020b} used a physical model similar to \cite{Fougere2016b} but instead of applying spherical harmonics to parameterise the AAF of the nucleus surface, they assumed a homogeneous nucleus composition (i.e., constant AAF).
The gas production rate was modulated by the illumination conditions only, i.e. there was no regional heterogeneity as in the works by \cite{Fougere2016b}, \cite{Combi2020}, \cite{Laeuter2020} but remained calibrated with Rosetta/ROSINA data as in the other studies.
They show, that matching daily averaged measurements is sufficient to estimate the global mass loss.
Reproducing the precise diurnal variation, with the associated regional heterogeneity, is not necessary to estimate the total mass loss.
All of the above mentioned approaches predict the same global mass loss within error bars.

This indicates that the shape and regional heterogeneity do not significantly contribute to variations in the production rate of 67P.
This is in line with findings from \cite{Marshall2019}, who have shown that on average 67P behaves almost like a spherical nucleus.
They also show that this is not true in general and different shapes and spin states can have a significant influence on the gas production rate.
Importantly, so long as the respective model preserves mass conservation at large distances it will correctly characterize the total gas production rate.

The rough global surface distributions found by \cite{Fougere2016b} and \cite{Combi2020} with the ``spherical harmonics model'', \cite{Laeuter2019} with the ``Haser+shape model'', and \cite{Zakharov2018} and \cite{Marschall2019} (the latter two both for the northern Hemisphere) with physical models are in agreement in the following sense.
There is, e.g., enhanced water emission from the northern hemisphere and CO$_2$ from the southern hemisphere of 67P during northern summer.
The water production rate follows to first order the sub-solar latitude. 
Though these are important first insights they are inadequate to link activity with surface morphology and evolutionary history including erosion, which should be the ultimate goal.
We should point out though that a diverse surface morphology might be the result of activity and not the driver of it.

Put another way, using a heuristic model, whichever it may be, instead of a physical model is sufficient to estimate the global production rate for 67P and other comets.
\cite{Combi2019Icar} used a semi-analytical model called the time-resolved model \citep[TRM;][]{Makinen2005Icar} to calculate global water production rates for 61 comets.
This illustrates the strength of such approaches to determine the global properties of comets.
A comet with a spin state and shape such that it behaves as a sphere \citep[in the sense described in][]{Marshall2019} should be similarly suitable for these heuristic models as 67P appears to be.

We hope to have convinced the reader of the usefulness of heuristic models in some cases but also shown that they are somewhat inadequate to link structures in the coma to emission distributions at the surface and through that to surface morphology.
This link requires physically consistent models, which we will discuss in the next section.
But, as mentioned above, even physically consistent models have spatial resolution limits and the appropriate error propagation needs to be accounted for \citep{Marschall2020a}.

\section{\textbf{State-of-the-art physical models}}\label{sec:stateOfTheArtModels}
The comet nucleus is the primary source of vapour and refractory particles in the coma (coma solids that emit gas and dust may constitute a secondary extended or distributed source of matter). 
The nucleus surface also acts as a boundary that may scatter or adsorb coma molecules. 
From a coma modelling perspective, the type of input information needed regarding the nucleus source depends on the coma model. 
Kinetic models based on the Boltzmann equation require the emission flux, temperature and velocity distribution function for each species specified at the nucleus/coma interface.
They also require the nucleus surface temperature when dealing with scattering or adsorption of coma molecules \citep{Bird1994,Bird2013}. 
Hydrodynamic models based on Euler (EE) or Navier-Stokes (NS) equations require the the flux, temperature, and drift speed on top of the Knudsen layer (the boundary between the non-equilibrium near-surface layer and the equilibrium fluid flow).
We refer an interested reader to \cite{Hirsch2007} and \cite{Rodionov2002P&SS} for more details on the theory of the fluid methods.
Thermophysical nucleus models, discussed in Section~\ref{sec_thermal}, provide outgassing rates, surface temperatures, and near-surface temperature gradients. 
This constitutes necessary, but not sufficient, information needed to calculate transmission velocity distribution functions, discussed in Section~\ref{sec_distr_func}, and Knudsen layer properties, mentioned in Section~\ref{sec_knudsen}. 
In general, the thermophusical model of the surface and gas environment model are coupled in both directions (i.e. they are interdependent).
In practice, though, these are treated independently.

\subsection{Boundary conditions}\label{sec:boundaryConditions}
\subsubsection{Thermophysical modeling of the nucleus} \label{sec_thermal}
A comet nucleus has a porous interior consisting of refractories, crystalline and/or amorphous water ice, and secondary highly volatile species such as $\mathrm{CO_2}$ and CO.
The vast majority of the surface is covered by an ice-free dust mantle that absorbs solar radiation and emits thermal infrared radiation. 
For 9P, 103P, and 67P very small exposed icy patches ($\mathrm{H_2O}$ and/or $\mathrm{CO_2}$) have been observed on the surface \citep{Sunshine2006Sci, Pommerol2015A&A, Raponi2016MNRAS, Fornasier2016Sci}.
State--of--the--art thermophysical models (e. g., Guilbert--Lepoutre et al., in this volume; \citealp{Groussin2007Icar, Rosenberg2010Icar, Groussin2013Icar, Davidsson2013Icar, Davidsson2021b, Herny2021, Marboeuf2014} and references therein) consist of the coupled differential equations for energy and mass conservation of the nucleus, that attempt to describe how such a system evolves as solar energy is transported by solid--state and radiative conduction, ice sublimates while consuming energy, vapour diffuses according to local temperature and pressure gradients while transporting energy by advection, and gas eventually escapes to space through the dust mantle or recondenses at depth while releasing latent energy. 
The solutions to these equations provide temperature, partial gas pressures, porosity, and abundances of solids as functions of latitude, time, and depth for the rotating and orbiting nucleus, as well as outgassing rates for each considered volatile. 

Numerical coma models are generally too computationally demanding to allow for the usage of a state--of--the--art thermophysical nucleus model. 
Therefore, simplified thermophysical models are employed that typically balance solar energy absorption, thermal re-radiation, and energy consumption by surface water ice sublimation \citep[e.g.,][]{Crifo2005, Zakharov2008,Marschall2019}. 
This simplification has at least three important consequences that may affect the accuracy of coma models. 

First, simplified nucleus models with surface ice become substantially cooler than realistic nucleus models with dust mantles (during strong sub--solar sublimation near perihelion, the former typically have surface temperatures of $\sim 200\,\mathrm{K}$, while the latter have $\sim 350\,\mathrm{K}$; \citealt{Groussin2007Icar}). 
This means that transmission velocity distributions are biased towards low molecular initial speeds, while initial translational temperatures and drift speeds are underestimated. 
This first issue could partially be mitigated by calculating the radiative equilibrium temperature, i.~e., omitting sublimation cooling altogether. 
This would approximately account for the gas heating taking place as it diffuses through the hot dust mantle on its way to the surface.  
Another time-efficient option is to apply lookup tables generated by more advanced thermophysical models. 
A rudimentary version of that approach was employed by, e.g.,  \cite{Tenishev2008}, \cite{Fougere2016a}, and \cite{Combi2020}.

Second, simplified models typically produce 1--2 orders of magnitude more gas compared to realistic models, for which a finite diffusivity of the dust mantle quenches the flow. 
This problem is typically handled by introducing an ``active area fraction'' that reduces the production rate (and thus the near-surface gas number density) to the observed level, by assuming that only 
parts of the surface is covered by ice. 
We will come back to this issue in Section~\ref{sec:acitveAreaFraction}.

Third, simplified nucleus models lack thermal inertia effects, caused by non--zero heat conductivity and heat capacity. 
Consequently, those models predict peak outgassing at local noon, while realistically modelled activity is strongest in the afternoon.
This makes the modelled coma too axis-symmetric about the Sun-comet line, at least for a spherical comet. 
Complex shapes, such as the one of comet 67P, introduce additional complexity in the pattern of activity.
Furthermore, nighttime activity artificially goes to zero, which typically is dealt with by setting a low but arbitrary background outgassing.
While a small nighttime activity is mostly a good approximation for $\mathrm{H_2O}$ it is not for more volatile species.
Nighttime activity of $\mathrm{CO_2}$ was observed both for 9P \citep{Feaga2007Icar} and 103P \citep{Feaga2014acm}.
Even for 67P, which is dominated by $\mathrm{H_2O}$, nighttime activity (likely of $\mathrm{CO_2}$) needed to be invoked to understand the dust coma dynamics \citep{Gerig2020Icar}.
Also at 67P the coma above the southern hemisphere showed strongly enhanced $\mathrm{CO_2}$/$\mathrm{H_2O}$ ratios during the poorly illuminated winter months early on in the Rosetta mission \citep{Hassig2015Sci}. 
More on this follows in section~\ref{sec:67P}.
These observed instances of nighttime activity need to be reflected in thermophysical modelling that goes into the boundary condition of coma models \citep[e.g.,][]{Pinzon2021}.
This third issue is interesting because it has so far not seemed to hinder the modelling of $\mathrm{H_2O}$.
This might indicate that water ice is very close to the surface thus making thermal inertia effects small enough that they cannot be picked up by coma models.
Or the effects are so nuanced that they have simply not been discovered yet.
Compared to $\mathrm{H_2O}$ thermal inertia effects need to be properly addressed with a thermal model that includes the thermal lag \citep{Pinzon2021}.
The measurements of $\mathrm{CO}$ \citep{Hassig2015Sci} on the southern winter hemisphere of 67P show even less diurnal variation and a more uniform outgassing pattern than even $\mathrm{CO_2}$.
This suggests that $\mathrm{CO}$ comes from even deeper layers.
Given the $\mathrm{CO}$ is much more volatile than $\mathrm{CO_2}$ this observation is not surprising.

\subsubsection{Transmission velocity distribution functions} \label{sec_distr_func}

Most kinetic coma models postulate the semi-Maxwellian velocity distribution function (SMVDF) \citep[e.g.,][]{HuebnerMarkiewicz2000} as a boundary condition.
To ensure that the gas has a semi-Maxwellian velocity distribution the initial expansion of the gas into the first cell needs to be taken into account.
It turns out that one cannot directly draw the velocity vectors for the molecules at the surface from an SMVDF because by the time they have expanded into the first cell above the surface their VDF has been altered.
Therefore, there is a need to define a transmission semi-Maxwellian velocity distribution (TSMVD) to draw from \citep[][]{HuebnerMarkiewicz2000}.
The SMVD and the TSMVD differ by a factor $cos(\theta)$, where $\theta$ is the angle between the emission direction and the surface normal. 
Using the TSMVD when drawing initial velocities at the surface will, as molecules with higher $v_z$  component ($v_z$ being the component of the velocity in the direction of the surface normal) start to overtake molecules with lower $v_z$  component, establish an SMVD distribution inside the volume.

If a SMVDF is the proper VDF can of course be debated.
\cite{SkorovRickman1995} calculated the velocity distribution of molecules emerging from a cylindrical channel with a sublimating floor. 
They found that the emerging distribution function was Maxwellian if the channel was isothermal, but noted strong deviations when a temperature gradient was present. 
\cite{DavidssonSkorov2004} used a Monte Carlo approach to study molecular migration within a granular medium, as well as the distribution function of molecules exiting the medium and entering an empty half--space. 
They too found that the distribution function was semi--Maxwellian for isothermal media. 
However, when the temperature falls with depth (typical mid-day conditions) the outflow is less collimated, and when the temperature increases with depth (typical of late afternoon and night) the outflow is more collimated, compared to the semi--Maxwellian. 
\cite{Liao2016} investigated the consequences of changing the degree of collimation for the global coma properties. 
They found that increased initial collimation tended to lower the number density and translational temperature and increase the drift speed. 
Reducing the collimation would presumably have the opposite effect. 

Therefore, a fully self--consistent nucleus/coma model requires a thermophysical model that provides near-surface temperature gradients and outflow rates, as well as a detailed kinetic model of the molecular velocity distribution function of emerging gas, that can be fed to the kinetic coma model. 
We should add though that we currently lack critical information on the details of the pores, such as their physical dimensions.

Coma molecules that impact the comet surface either scatter or are adsorbed. 
Scattering occurs either through specular reflection or diffusively (in the latter case, thermalization on the nucleus surface may modify the molecular speeds). 
Adsorption gives rise to a (sub)--monolayer of volatiles on top of the dust mantle, that may have a short residence time. 
Because of the low surface density, their desorption rates should ideally be calculated with the first--order Polanyi--Wigner equation \citep[e.g.,][]{Suhasaria2017}, using an activation energy that is suitable for vapour molecules attached to a silicate or organics surface. 
This is different from the zeroth--order sublimation that typically is considered for multilayer ice deposits. 
Scattered and thermally desorbed molecules form a separate population that should be added to obtain the complete transmission velocity distribution function.
We have ample evidence that this population exists and can be adsorbed from the ambient coma \citep{Liao2018} or accumulate when the interior is warmer than the surface \citep{DeSanctis2015}.

\subsubsection{The Knudsen layer} \label{sec_knudsen}
The quasi--semi--Maxwellian velocity distribution emerging from a sublimating medium relaxes to a drifting Maxwell--Boltzmann velocity distribution over a finite distance due to molecular collisions.
Within this ``Knudsen layer'' \citep[e.g.,][]{Cercignani2000} the gas properties evolve according to 
the Boltzmann equation with a non--zero collision integral. 
At the upper boundary of the Knudsen layer (if it has a finite thickness), the collision integral goes to zero, so that the first three moments of the Boltzmann equation approaches the NS and, further, the Euler equations, i.e., a hydrodynamic formulation becomes valid in the downstream flow.

The gas number density, translational temperature, and drift speed at the boundary (i.e., the top of the Knudsen layer) are connected to the nucleus surface temperature and near--nucleus number density through the ``jump conditions''. 
These were defined by \cite{Anisimov1968}, who assumed Mach number $M=1$ at the upper boundary, while \cite{Ytrehus1977} demonstrated that the problem does not form a closed set of equations so that the solution becomes a function of an assumed downstream Mach number. 
As described by \cite{Crifo1987}, the boundary conditions and the downstream hydrodynamic solutions, therefore, need to be brought in agreement through an iterative procedure. 
For further discussions about cometary Knudsen layers and application of jump conditions, see e.~g. \cite{Davidsson2008} and \cite{Davidsson2021Icar}. 
If the Knudsen layer is thin, the jump conditions allow for the specification of inner coma boundary conditions in hydrodynamic coma models. 
Comparisons between the kinetic and hydrodynamic versions of coma models have been made by, e.~g., \cite{Crifo2002bNSDSMC, Crifo2003} and \cite{Zakharov2008}.

\subsubsection{Active area fraction}\label{sec:acitveAreaFraction}
As described above, instead of a full thermophysical model, a simplified thermal balance is often employed to determine the boundary conditions for the dynamical models.
This leads to the introduction of an ``active area fraction'' (AAF) to reduce the flux, usually one to two orders of magnitude too high, to realistic values.
The AAF is simply a linear term multiplied with the gas production rate.

In recent years this AAF, though called differently by different groups, has been the main parameter used to fit the observed number/column densities \citep[e.g.,][]{Fougere2016a,Marschall2016,Zakharov2018}.
Naturally, the question arises as to the physical interpretation of the found AAFs.
Given that they are based on exposed pure ice surfaces, which we do not observe apart from very few isolated patches \citep[e.g.,][]{Sunshine2006Sci,Pommerol2015A&A}, the absolute values of the AAF should not be taken literally (i.e., having AAF~$=0.05$
does not mean that 95\% of the surface is dust and 5\% is literally exposed water ice).
If that were true then and instrument such as Rosetta's VIRTIS would have seen $3~\mu$m water ice absorption over vast swaths of the surface.
This was not the case on comet 67P where VIRTIS saw the absorption only at a few specific locations \citep[e.g.,][]{Barucci2016A&A,Raponi2016MNRAS}.
The relative values of AAF reveal actual differences in the response of the surface to solar illumination.
The cause of these differences is not captured in the AAF.
But, apart from isolated patches of water ice the surface is covered by a dry layer.
This layer quenches the flow from the sublimation front beneath it and thus provides a better stand in for understanding the AAF.

Though the initial surface number density, temperature and velocity are the common input parameters for gas kinetic models, we have seen that the simplified model leads to surfaces that are significantly cooler than realistic nucleus models with dust mantles and thus lead to slower initial molecular speed.
The scaling through AAF ensures that the total flux is of the correct order of magnitude but the temperatures and speed of the gas flows may not be.
Observations of the gas number/column density are rather insensitive to this but measurements by Rosetta/MIRO indicate that deviations from the model temperatures and velocities are observed \citep{Marschall2019}.
In essence, the model production rates are to a large extent solid but the gas speeds and temperatures are not.

The AAF is nevertheless useful because it highlights real differences in how ``active'' different regions are.
Presenting only local gas production rates would make it almost impossible to determine if two respective regions show different levels of activity because of different illumination conditions or because of, e.g., different ice contents or variations in the dust cover.
The AAF attempts to remove any diurnal and seasonal variations caused by the local illumination conditions.

The AAF is an important quantity to parameterise the asymmetry of surface activity but can be problematic if used to fit data far from a comet or of comets which have a significant secondary contribution of gas from icy grains in the coma (extended source, see Sec.~\ref{sec:extendedSource} for more).
In the case of an extended gas source, the AAF will combine the effects of surface and coma sublimation terms and is no longer representative of the surface property.


\subsection{\textbf{Kinetic and fluid models}} \label{sec:kineticDynamicModels}

\begin{figure*}
	\includegraphics[width=\textwidth]{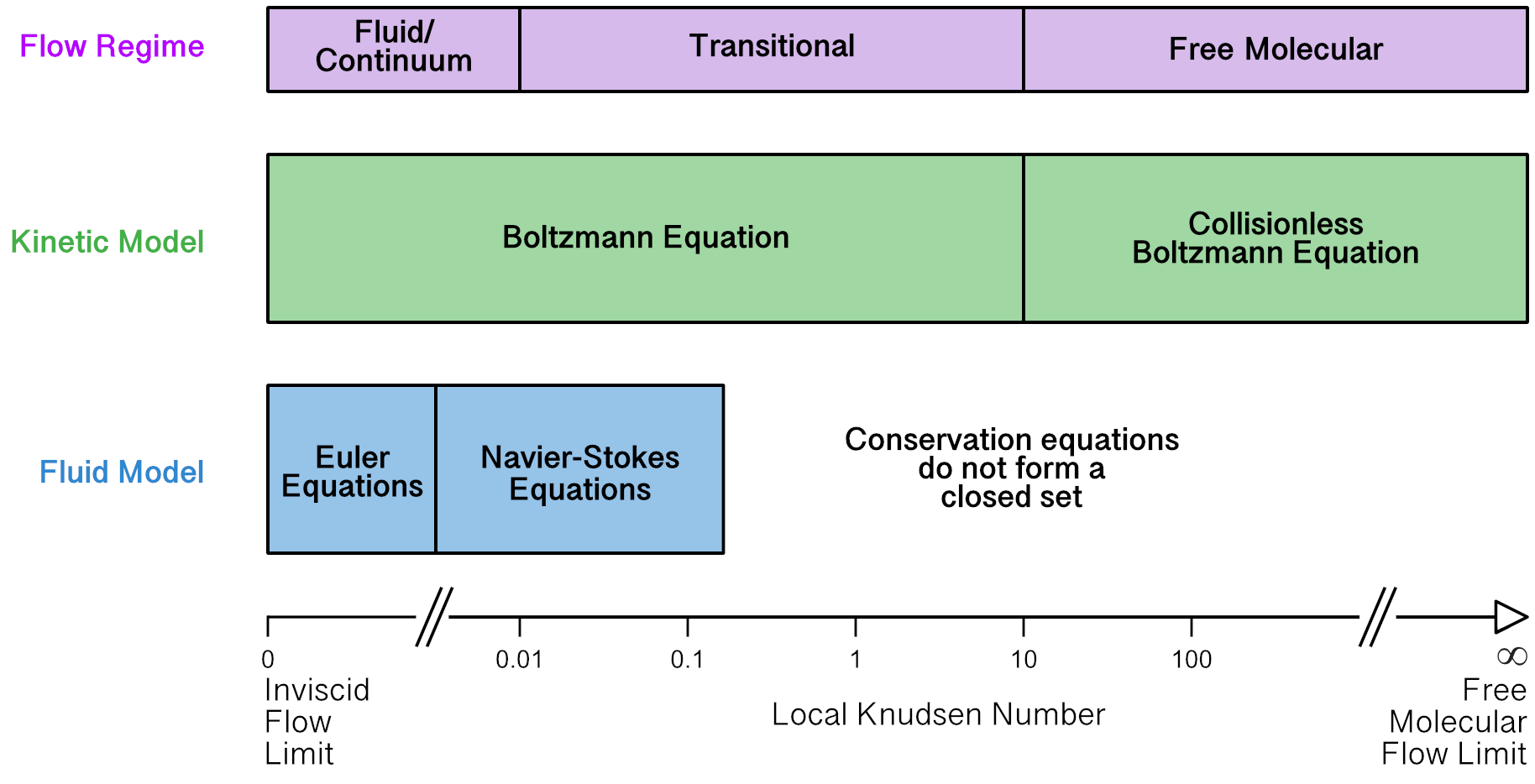}
	\caption{Different flow regimes as a function of the Knudsen number as well as the intervals for which different models (fluid and kinetic) are valid. This figure has been adapted from its original in \cite{Bird1994}.}
	\label{fig:flowRegimes} 
\end{figure*}

In most cases of practical interest, studying cometary comae involves the consideration of rarefied gas flows under strong non-equilibrium conditions.

Kinetic modelling of cometary comae is based on solving the Boltzmann equation (Fig.~\ref{fig:flowRegimes}):
\begin{equation}
    \frac{\partial f}{\partial t}+\mathbf v \frac{\partial f}{\partial\mathbf x} + \dot{\mathbf v} \frac{\partial f}{\partial\mathbf v}={I}(\mathbf v),
    \label{boltzmann-eq}
\end{equation}
where $f=f(\vec x, \vec v, t)$ is a distribution function of atoms or molecules of the gas in phase space where $\vec x$, $\vec v$, $t$ are the coordinate, velocity and time respectively.
$I(\mathbf{v})$ is a collision integral describing the interaction of these particles.

Solving the Boltzmann equation is a challenging problem that is further complicated by the need to account for the complexity of the nucleus shape. 
Typically, kinetic models developed within the Direct Monte Carlo Method (DSMC) method are used for kinetic modelling of the dusty gas dynamics in cometary comae. 
This modeling approach allows to include important processes such as inelastic inter-molecular collisions, photochemical reactions, and interaction with dust \citep[e.g., ][]{Combi1996Icar,Tenishev2008,Tenishev-2021-JGR,Crifo2003,Crifo2005,Zakharov2008}. 
Compared to fluid models the main advantage of a kinetic model is that it is valid at any degree of non-equilibrium and/or rarefaction (i.e. conditions typical in the coma; Fig.~\ref{fig:flowRegimes}).
Another approach to model gas dynamics in cometary comae is based on solving the Navier–Stokes equation -- a fluid model (Fig.~\ref{fig:flowRegimes}) -- as detailed by e.g., \cite{Crifo2002bNSDSMC}.

The Rosetta mission delivered a large volume of new data.
This data includes, a.o., almost continuous monitoring over two years of i) the gas density and composition at the location of the spacecraft (ROSINA), ii) spectral imaging of the gas coma (VIRTIS), coma LOS spectra (VIRTIS, MIRO), and imaging of the nucleus and dust coma by the `Optical, Spectroscopic, and Infrared Remote Imaging System' \citep[OSIRIS,][]{Keller2007SSRv}.
Kinetic models based on the DSMC method demonstrated robustness when they were applied for interpretation of these new data \citep[e.g., ][]{Fougere2016a,Marschall2016,Liao2016}.

In the most general case, the flow in the coma covers regions with widely differing conditions – from fully collision-less to fluid. 
The degree of rarefaction is characterized by the Knudsen number, $Kn$:

\begin{equation}\label{eq:Kn}
Kn=\lambda/L\quad,
\end{equation}

where $\lambda$ is the MFP of the molecules and $L$ is a characteristic length. 
The choice of $L$ is not immediately clear and depends on the problem under consideration.
When describing the entire flow within the near nucleus environment by a single global Knudsen number the equivalent radius of the nucleus is traditionally used for $L$ as the characteristic scale of the flow. 
When characterising the rarefaction of the flow locally, the scale length of the macroscopic gradient can be used as $L$ (e.g., using the gas density, $n_g$, such that $L = n_g/|\nabla n_g|$). 
Depending on the local $Kn$ three flow regimes can be roughly distinguished (illustrated in Fig.~\ref{fig:flowRegimes}): 

\begin{enumerate}
    \item continuum/fluid, $Kn<0.01$;
    \item transitional, $0.01 \leq Kn \leq 100$;
    \item free molecular, $Kn>100$.
\end{enumerate}

The expansion of the flow leads to a decrease in collisions with radial distance and therefore, even if an equilibrium flow was established at the top of the Knudsen layer, the flow becomes non-equilibrium again at larger distances due to an insufficient collision rate to maintain the equilibrium distribution.

For a more detailed description of the numerical methods, we refer the reader to \cite{Hirsch2007} and \cite{Bird1994}, as well as \cite{Bird2013} for the DSMC method specifically.
For a recent deeper description of the different models as well as a historical review of their development we refer to \cite{Marschall2020c}.

\subsection{Flow regime estimations}\label{sec:regimeEstimates} 
To select the appropriate approach (fluid or kinetic) we need to know which flow regime our cometary coma covers.
To get a more quantitative understanding of the scales involved and which flow regimes can be expected, a simple order of magnitude estimation can be made using Eq.~\ref{eq:Haser} and the fact that the MFP is

\begin{equation} \label{eq:mfp}
\begin{split}
    \lambda &= \frac{1}{\sqrt{2} n \sigma} \\
            &= \frac{4\pi r^2v}{\sqrt{2} Q\sigma}                    \quad,
\end{split}
\end{equation}
where $\sigma$ is the collisional cross-section of the molecules.
The MFP thus scales with the square of the distance to the center of the nucleus but only linearly with the gas speed and inversely proportional to the gas production rate.

\begin{figure*}
	\includegraphics[width=\textwidth]{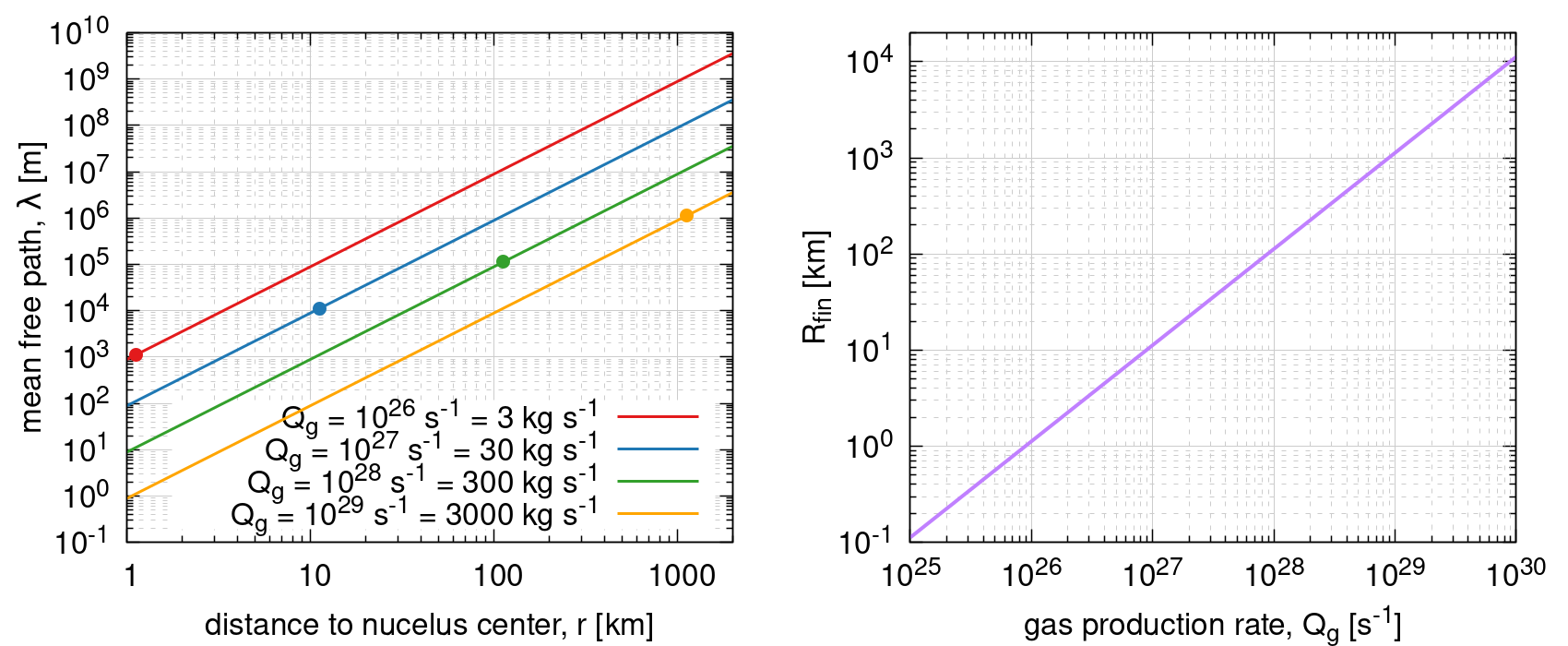}
	\caption{The left panel shows the mean free path, $\lambda$, of H$_2$O as a function of the distance to the centre of the nucleus for four different gas production rates spanning four orders of magnitude according to Eq.~\ref{eq:mfp}. The coma is assumed to be isotropic and expanding at a constant speed. Here we assume $v=1000$~m/s and $\sigma_{H_2O}=10^{-19}$~m$^2$. The dots indicate the distances beyond which we expect the last collision. 
	The right panel shows the distance, $R_{fin}$, as a function of the gas production rate. $R_{fin}$ is the distance beyond which we expect the last collision of a molecule with another. The respective expression is given in full in Eq.~\ref{eq:Rfin}. All other assumptions are the same as in the left panel.}
	\label{fig:mfp} 
\end{figure*}

The left panel of Figure~\ref{fig:mfp} shows the MFP as a function of the distance to the centre of the nucleus.
As a reminder, for Eq.~\ref{eq:mfp} to be valid we assume an isotropic coma which expands at a constant speed.
Here we have set the expansion speed, $v$, to $1000$~m/s.
In reality, $v$ is not independent of the gas production rate but $1000$~m/s represents a reasonably realistic speed for the order of magnitude estimation we perform here.
In general, the collisional cross-section, $\sigma$, is a function of temperature (or the relative velocity of colliding particles). 
It is constant only in the hard sphere model.
For simplicity we assume $\sigma_{H_2O}=10^{-19}$~m$^2$.

For very low water production rates ($10^{26}$~molecules/s = $3$~kg/s) the MFP is large even close to the nucleus.
It quickly expands from 1~km at 1~km from the nucleus centre to 10,000~km at 100~km from the nucleus centre (left panel of Figure~\ref{fig:mfp}).
Even for the highest water activity case ($10^{29}$~molecules/s = $3,000$~kg/s) the MFP is of the order of 10~km at a distance of 100~km from the nucleus centre.

As a reference the nucleus of comet 67P had an water production rate of $\sim10^{26}$~molecules/s = $3$~kg/s at a heliocentric distance of 3~au inbound a few months after Rosetta arrived \citep{Marschall2020b,Combi2020}.
67P reached a peak production rate of the order of $1,000$~kg/s shortly after perihelion at a heliocentric distance of $1.25$~au \citep[estimates range between 500 and 1,500 kg/s, depending on the instrument used and given that $\mathrm{H_2O}$ accounts for  80\% of the total volatile mass loss;][]{Hansen2016MNRAS, Fougere2016b, Marshall2017A&A, Kramer2017, Combi2020,Marschall2020b}.
During the 1P/Halley encounters in situ measurements by \cite{Krankowsky1986} showed a total gas production rate of 
6.9 $\times$ 10$^{29}$ molecules/s $\approx20,000$~kg/s.

Another useful concept is the distance beyond which we expect the final collision.
We shall call this distance $R_{fin}$ and can find it by

\begin{equation}
   1 \doteq \int_{R_{fin}}^{\infty} \frac{dr}{\lambda} = \frac{\sqrt{2}Q\sigma}{4\pi v R_{fin}}\\
\end{equation}
and hence
\begin{equation} \label{eq:Rfin}
    R_{fin} = \frac{\sqrt{2}Q\sigma}{4\pi v} \quad.
\end{equation}
Thus, for the isotropic coma expanding at constant speed the distance beyond which we expect only one last collision is proportional to the production rate and inversely proportional to the expansion speed of the gas. 
Of course, this is a big simplification.
The assumption that -- after its initial expansion -- the gas speed is constant is a fairly good one for typical comets up to a distance of at least $10^4$~km when considering H$_2$O \citep[see, e.g., Fig.~2 in][]{Shou2016ApJ}.
The distance at which the gas starts to accelerate again is of the order of $10^4$~km and is the result of a selection effect where dissociation and ionization preferentially removes the slower molecules from the phase space \citep{Tseng2007A&A}.
Therefore, this is a fairly good order of magnitude estimate.
The right panel of Figure~\ref{fig:mfp} shows $R_{fin}$ as a function of the gas production rate.
For very low water production rates ($10^{26}$~molecules/s = $3$~kg/s) $R_{fin}$ is only one kilometer.
This means that for a nucleus larger than one kilometre and with such a low water production rate the molecules will only experience one collision.
The flow is essentially in the free molecular regime from the surface on.
For the higher water activity cases (e.g., $10^{29}$~molecules/s = $3,000$~kg/s) the last collision can only be expected beyond 1,000~km.

\begin{figure*}
	\includegraphics[width=\textwidth]{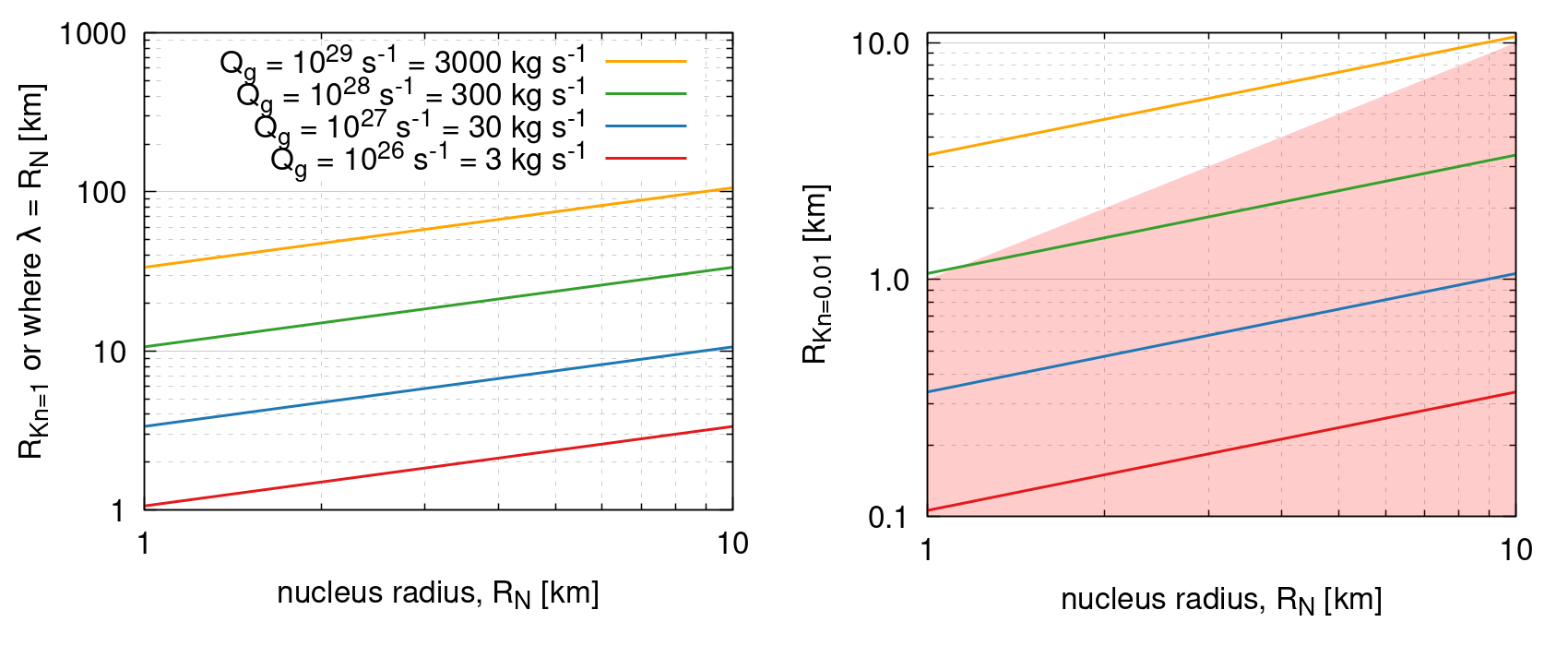}
	\caption{The left panel shows the distance at which the mean free path, $\lambda$, is equal to the radius of the nucleus, $R_N$, for four different production rates spanning four orders of magnitude and as a function of the $R_N$. The other assumptions are as in Fig.~\ref{fig:mfp}. This distance is at the location where $Kn=1$. Therefore, the lines shown are given by Eq.~\ref{eq:RKn} when $Kn=1$.
	The right panel shows the same as the left panel but for $Kn=0.01$ in Eq.~\ref{eq:RKn}. The value of $Kn=0.01$ corresponds to the transition between the fluid and transitional regimes. The red shaded area denotes distances to the centre of the nucleus which lies below the current nucleus surface.}
	\label{fig:locationOfKn} 
\end{figure*}

Further, we would like to know at which distance we transition from one flow regime to another (for example from the fluid to the transitional).
If we define the distance $R_{Kn}$ where the Knudsen number is defined with $L=R_{Kn}$.
We can thus combine Eq.~\ref{eq:Kn} and Eq.~\ref{eq:mfp} and find that
\begin{equation} \label{eq:RKn}
   R_{Kn} = \sqrt{\frac{\sqrt{2}Q\sigma R_N Kn}{4\pi v}} \quad.
\end{equation}

Two cases of this equation are particularly interesting.
First, at which point is the MFP as large as the radius of the nucleus, $R_N$?
It turns out that this is equivalent to finding $R_{Kn}$ for $Kn=1$.
This means that when we model a coma environment that has the size $R_{Kn=1}$ then the global Knudsen number is unity and we are squarely in the transitional flow regime and therefore likely need to apply DSMC to capture the physics correctly.
The left panel of Figure~\ref{fig:locationOfKn} shows $R_{Kn=1}$ for different nucleus radii between 1 and 10~km and four gas production rates spanning four orders of magnitude.
Any domain larger than the given $R_{Kn=1}$ will have a global $Kn\ge1$.

Second, we would like to find the approximate location of flow transitions.
For example, we'd like to know when our system is between the fluid and transitional regime, i.e. the regime that can be safely studied using the more efficient fluid solvers (Euler/NS) compared to the regime which will require a kinetic approach (DSMC).
To be conservative this transition occurs around $Kn=0.01$.
Thus, the right panel of Figure~\ref{fig:locationOfKn} shows $R_{Kn=0.01}$.
Note that the shaded red area marks the interior of the nucleus up to the surface.
For all production rates below $Q=10^{28}$ ($\sim 300$~kg/s) the respective lines fall below the size of the nucleus.
This means that the flow will be in the transitional regime above the surface for any nucleus larger than 1~km.
Simulation domains that are larger than $R_{Kn=0.01}$ will have a $Kn\ge0.01$, i.e. the global regime shifts further towards free molecular flow.
Only for much higher production rates, and smaller nuclei, will we find several meters to kilometres of fluid regions above the surface.
When one considers a larger $Kn$ as the transition between fluid and transitional flow, e.g., $0.1$ instead of $0.01$, then the respective $R_{Kn}$ distances increase by a factor of $\sqrt{10}\sim3$.
Therefore, for all production rates below $Q=10^{27}$ ($\sim 30$~kg/s) the transition from fluid to transitional flow happens "below" the nucleus surface.

Note also that in Eq.~\ref{eq:RKn} all terms ($R_N$, $Q$, $Kn$, and $v$) enter with the same weight.
For the same $Kn$, any increase of the nucleus radius is compensated by the respective decrease in gas production rate (see the right panel in Fig.~\ref{fig:locationOfKn}).

\subsection{\textbf{Dusty-gas flow}}\label{sec:dustyGasFlow}
While we have focused here on the gas flow and, to a large extent, neglected the presence of dust particles in the flow, we would be remiss not to discuss some fundamental aspects of dusty-gas flows.
For the most part, dust particles are treated as passive objects that act as test particles within the gas flow.
Of course, this is not true in general.
For a detailed discussion of the dust dynamics, we refer the reader to Agarwal et al. in this volume.

In the most ideal and straightforward case, we are dealing with a coma containing dry dust and a dust-to-gas mass flux ratio much smaller than one (i.e., low dust content).
In this case -- where the back-coupling from the dust to the gas can be neglected -- the gas flow can be treated independently from the dust flow.
Separate models for the gas and dust flows can be run sequentially.
This not only allows for more flexibility to explore the parameter space but is also much less computationally expensive because of the different time scales of gas molecule and dust particle motion.
It is thus a convenient scheme to employ.

Here we will highlight how the presence of the dust can alter the gas flow, thus deviating from this ideal case and thus requiring a coupled dust-gas coma model.

\subsubsection{Momentum transfer}\label{sec:momentum-tranfer}
If the dusty-gas flow has a high dust-to-gas mass ratio the two flows cannot be treated independently. 
When the dust particles are accelerated by the gas flow, momentum is transferred from the gas to the dust.
As the dust-to-gas ratio increases so do the kinetic energy and total momentum of the dust flow.
When momentum transferred from gas to dust becomes a significant fraction of the gas flow momentum, the presence of dust becomes noticeable for the gas flow as well and the gas flow will be slowed down.
There is an important caveat though.
If the dust is hotter than the gas, it will be able to accelerate the molecules it interacts with (see Sec.~\ref{sec:energy-transfer}).

Whether or not the dust flow impedes the expansion of the gas coma depends not only on the dust-to-gas ratio but also on the dust size distribution.
A coma with large slow-moving particles will -- for a given total dust mass -- impact the gas flow significantly less than the case where the same mass is distributed in small particles that accelerate to a significant fraction of the gas speed.
As far as we are aware, no study has quantified the combined effect of the dust size distribution and dust-to-gas ratio but the reader shall be aware of this potential pitfall.


\subsubsection{Energy transfer}\label{sec:energy-transfer}
A further effect of the presence of dust particles in the gas flow occurs from the fact that dust particles can be significantly hotter than the surrounding gas.
The gas cools as it expands into space while dust particles heat up when exposed to the Sun \citep{Lien1990}.
This makes the temperature gap between gas and dust larger with increasing cometocentric distances.
Additionally, sufficient molecule-dust collisions need to occur which typically happens close to the surface.
Therefore, collisions of the gas molecules with the dust particles will increase the gas temperature.
The presence of small (micron and sub-micron) particles can increase the gas temperature by as much as a factor of 3 \citep{Kitamura1987,Markelov2006}.
A mass-loaded gas flow will thus initially be slowed down and then subsequently heated in the inner part of the coma \citep{Crifo2002a}.
Radiative heating of the gas by the hotter dust can become a dominant effect.

As in the case of momentum transfer (Sec.~\ref{sec:momentum-tranfer}) the effect of energy transfer becomes of particular concern when the coma is dominated by very small particles, which are more easily heated to very high temperatures.
Super-heated dust particles have been observed at various comets \citep[e.g., at 1P, Hale-Bopp, and 67P;][]{GehrzNey1992, Williams1997, BockeleeMorvan2019}.

Though both momentum transfer and heating from dust can significantly alter the gas flow, we currently don't know the extent to which such particles have altered an observed gas flow.
No observational constraints are currently available to evaluate the effect of momentum and heat transfer within the gas flow due to mass loading.

\begin{figure}[ht!]
\begin{center}
\includegraphics[width=0.32\textwidth]{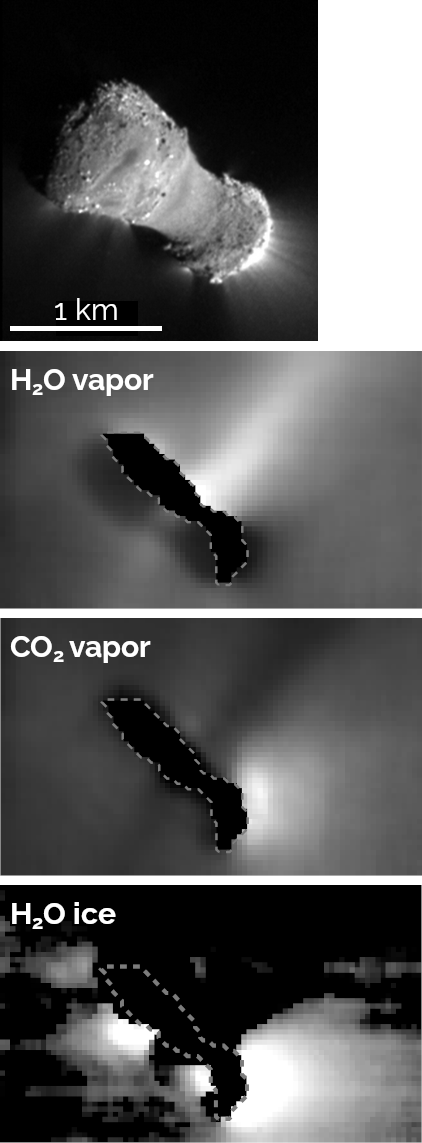}
\caption{Top panel shows the nucleus of 103P/Hartley 2 taken with the MRI instrument for context. The two panels labelled ``H$_2$O/CO$_2$ vapor'' are maps of the total flux in the relevant emission bands. The bottom panel labelled ``H$_2$O ice'' shows a map of the depth of the water ice absorption feature at $3\mu$m. The three lower panels were taken 7 minutes after the closest approach. The figure was adapted from data presented in \cite{AHearn2011}\protect\footnotemark\ for the top panel and \cite{Protopapa2014}\protect\footnotemark\ for the lower three panels.
}
\label{fig:hartley2-EPOXI}
\end{center}
\end{figure}

\addtocounter{footnote}{-1}
\footnotetext[\thefootnote]{Courtesy NASA/JPL-Caltech; \url{https://photojournal.jpl.nasa.gov/jpeg/PIA13570.jpg}}
\addtocounter{footnote}{1}
\footnotetext[\thefootnote]{Credit: Courtesy of Silvia Protopapa through private communications. The panels represent composition maps of the coma of comet 103P/Hartley~2 published by \cite{Protopapa2014} and derived from data publicly available through the Small Bodies Node (SBN) of NASA's Planetary Data System (PDS) \citep{McLaughlin2013}.}

\subsubsection{Mass transfer}\label{sec:extendedSource}
Another back-coupling from the dust to the gas comes from `wet' dust.
When dust particles consist of ice or contain a large fraction of ice, they will begin to sublimate and contribute to the gas coma in the form of an extended/distributed gas source.
The sublimation of icy or ice-rich dust particles directly alters the dynamics of the dust particles \citep{Kelley2013,Agarwal2016,Guettler2017}.

The contribution from such icy particles can be significant and it might explain hyperactive comets \citep{Sunshine2021}.
But a recent non-detection of icy grains in the coma of hyperactive comet 46P/Wirtanen by \cite{Protopapa2021} challenges this interpretation of hyperactivity.

Hyperactive comet 103P/Hartley 2 -- visited by the extended Deep Impact mission EPOXI \citep{AHearn2011} -- shows an abundance of icy particles (bottom panel in Fig.~\ref{fig:hartley2-EPOXI}; \citealp{Protopapa2014}).
The icy grains, driven by CO$_2$ activity (bottom two panels in Fig.~\ref{fig:hartley2-EPOXI}), sublimate in the coma while larger chunks can redeposit in the neck region where they could cause the observed water ``jet'' ($2^{nd}$ panel from the top in Fig.~\ref{fig:hartley2-EPOXI}).

Sublimating icy grains lead to a slower expansion but warmer/hotter gas compared to a comet nucleus source only. 
This has been observed and modelled for comet 73P/Schwassmann-Wachmann 3 \citep{Fougere2012} and Hartley 2 \citep{Fougere2013}.

\subsubsection{Modeling dusty gas flows}
The first multidimensional models (axially-symmetric and 3D) of a dusty gas coma with the physically consistent description of a flow as a fluid were presented in \cite{Kitamura1986,Kitamura1987,Kitamura1990}, and \cite{KorosmezeyGombosi1990}. 
These models were based on the numerical solution of the coupled hydrodynamic equations (representing mass, momentum, and energy conservation). 
The dust was treated as one of the components of the fluid consisting of single-sized spherical grains ($<1 \mu m$). 

However, aspherical grains affect the maximum liftable sizes and velocity distribution. 
Unlike spherical particles which experience only drag force, aspherical particles have also transversal (i.e., lift) force and torques \citep[see][]{Ivanovski2017}.
As a consequence they may start to rotate thus altering the trajectories with respect to their spherical counterparts.
Aspherical particles may also have rotational energy and therefore act as an additional sink of energy from the gas.
Agarwal et al. in this volume go into more detail about dust dynamics.

Modern numerical models of the dusty gas flow \citep[e.g.,][]{Tenishev2011} were constructed in the spirit of DSMC for the gas.
The dust phase in the coma is represented by a large but finite number of model particles that represent real dust grains. 
The motion of a spherical, isothermal particle in the inertial frame is assumed to be governed by the gas drag and nucleus gravity force:
\begin{equation}\label{eq:dust-motion-eq}
    \frac{4\pi}{3}\rho_d a^3\frac{d\mathbf v_d}{dt}=\frac{1}{2}a^2C_D \rho\left|\mathbf u-\mathbf v_d\right|\left(\mathbf u-\mathbf v_d\right)-\mathbf{F}_g \quad.
\end{equation}
Here, $v_d$ is the velocity of a spherical dust particle with radius $a$ and bulk density $\rho_d$, $C_D=C_D(\mathbf u, \mathbf v_d, T_g, T_d)$ is the drag coefficient which depends on the gas and dust flow parameters, $F_g$ the gravity force, $u$ and $\rho$ the gas velocity and density, respectively.

The drag coefficient is a function of relative velocity of the gas and dust and of their temperatures. 
Typically the MFP in the coma is much larger than the size of the dust grains (meters vs. mm -- microns) therefore $C_D$ can be taken for a particle in a free molecular flow \citep{Bird1994}.
Because the drag coefficient asymptotically approach 2 for most dust particle sizes, $C_D =2$ is often assumed rather than the more complicated and complete description.
The choice of $C_D =2$ implies, though, that the acceleration of particles will in general be underestimated (for small particles it’s strongly underestimated).
See Agarwal et al. in this volume for more details on this topic.

The dynamics of the dust grains are significantly affected by the geometry and mass distribution of the nucleus. 
The complexity of the nucleus surface geometry, in turn, complicates the calculation of the gravity force acting upon a dust particle.
Realistic gravity fields must therefore be incorporated into the dynamical models \citep{Tenishev-2016-MNRAS,Marschall2016}.

In the vicinity of the nucleus (distances of a few tens of nucleus radii) additional forces such as solar radiation pressure and solar gravity can be neglected and are therefore absent in Eq.~\ref{eq:dust-motion-eq}.
Additionally, Eq.~\ref{eq:dust-motion-eq} assumes spherical particles and thus does not include terms related to the rotation of particles \citep{Ivanovski2017}.

The distinctive feature of a model, like the one presented in \cite{Tenishev2011}, is the self-consistent kinetic treatment of both the gas and the dust phases of the coma. 
These numerical experiments with that model suggested that the effect of dust on the gas flow in the coma is minimal. 
Hence, in modelling the dust phase of the two-phase dusty gas from in the coma, the effect of the dust phase on the gas can in most cases be neglected.
However, we should add that this statement should be checked for a wider range of parameters in the future.

Importantly, within a region of roughly $10-100~R_N$ the assumption that the gas flow is in steady state can generally be used.
In contrast, this is not true for the dust -- at least not for all dust sizes.
For a more precises estimate of the steady state distance we need to take into account the rotation period of the nucleus, $T_{rot}$, which modulates the surface emission distribution (i.e., the boundary condition of the flow).
There is no absolute threshold but let us suppose the nucleus rotation can be neglected, e.g., when it rotates by less than $5^{\circ}$.
In this case the flow is steady on the length scale of $L_{S}=vT_{rot}\frac{5^{\circ}}{360^{\circ}}$.
For a nucleus with a rotation period of 12~hours and a water expansion speed of $v_{gas}=1,000$~m/s we get $L_S = 600$~km.
This is a much shorter scale than, e.g., the scale for photo-dissociation.
For slow dust particles moving at, e.g., $v_{dust}\sim10$~m/s we get $L_S = 6$~km.
For typical nuclei sizes this is within the typical dust acceleration region of $10R_N$.

\begin{figure*}[ht!]
\begin{center}
\includegraphics[width=0.89\textwidth]{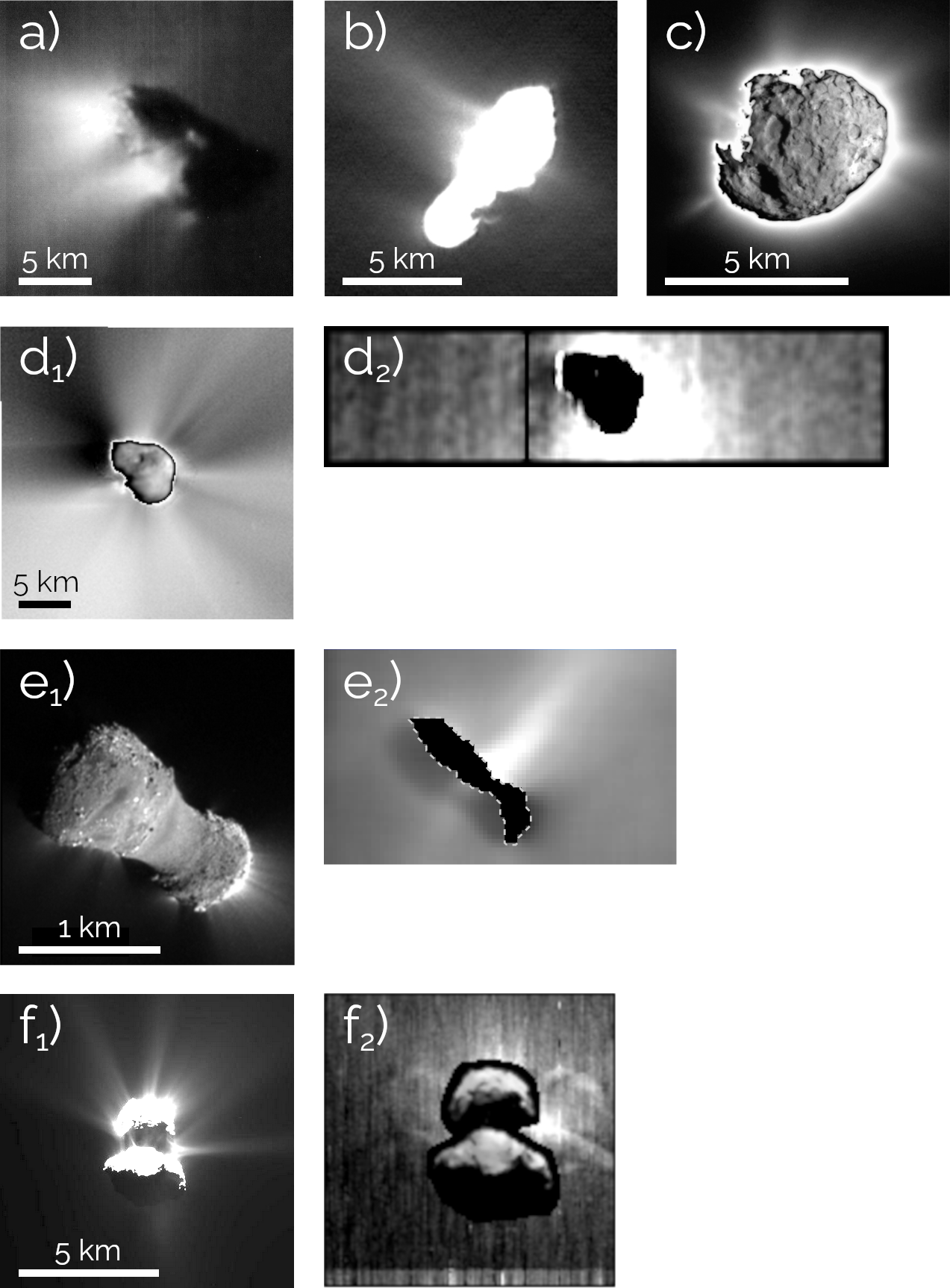}
\caption{In chronological order of comets visited: The inner dust structures of comet 1P/Halley \citep[panel a;][]{Keller1987}\protect\footnotemark, 19P/Borrelly \citep[panel b;][]{Boice2002}\protect\footnotemark, 81P/Wild 2 \citep[panel c;][]{Tsou2004}\protect\footnotemark, 9P/Tempel 1 \citep[panel d$_1$;][]{Feaga2007Icar}\protect\footnotemark, 103P/Hartley 2 \citep[panel e$_1$;][]{AHearn2011}\protect\footnotemark, and 67P/Churyumov–Gerasimenko \citep[panel f$_1$;][]{Tubiana2019}\protect\footnotemark. Maps of the water column density are shown for the three most recently visited comets by the Deep Impact/EPOXI and Rosetta missions: 9P \citep[panel d$_2$;][]{Feaga2007Icar}\protect\footnotemark, 103P \citep[panel e$_2$;][]{Protopapa2014}\protect\footnotemark, and 67P (panel f$_2$)\protect\footnotemark, respectively.
}
\label{fig:gasAndDustStructures}
\end{center}
\end{figure*}

\addtocounter{footnote}{-8}
\footnotetext[\thefootnote]{Credit: ESA/MPS under Creative Commons Attribution-Share Alike 3.0 igo (\url{https://creativecommons.org/licenses/by-sa/3.0/igo/deed.en}).}
\addtocounter{footnote}{1}
\footnotetext[\thefootnote]{Courtesy NASA/JPL-Caltech (\url{https://photojournal.jpl.nasa.gov/catalog/PIA03501}).}
\addtocounter{footnote}{1}
\footnotetext[\thefootnote]{Courtesy NASA/JPL-Caltech (\url{https://photojournal.jpl.nasa.gov/catalog/PIA05578}).}
\addtocounter{footnote}{1}
\footnotetext[\thefootnote]{Adapted from \cite{AHearn2008EP&S} under a Creative Commons Attribution 4.0 International License (\url{https://creativecommons.org/licenses/by/4.0/})}
\addtocounter{footnote}{1}
\footnotetext[\thefootnote]{Courtesy NASA/JPL-Caltech (\url{https://photojournal.jpl.nasa.gov/jpeg/PIA13570.jpg})}
\addtocounter{footnote}{1}
\footnotetext[\thefootnote]{Rosetta/{\allowbreak}OSIRIS/{\allowbreak}WAC image taken on 2015-04-27 18.17.57.683 UTC; Credit: ESA/{\allowbreak}Rosetta/{\allowbreak}MPS for OSIRIS Team MPS/{\allowbreak}UPD/{\allowbreak}LAM/{\allowbreak}IAA/{\allowbreak}SSO/{\allowbreak}INTA/{\allowbreak}UPM/{\allowbreak}DASP/{\allowbreak}IDA published under the Creative Commons license CC BY-SA 4.0.}
\addtocounter{footnote}{1}
\footnotetext[\thefootnote]{Credit: Courtesy of Lori Feaga through private communication. The panel represents a map derived from the current state of calibration of the data first published by \cite{Feaga2007Icar} for 9P/Tempel~1. The data from which the map is derived is publicly available through the Small Bodies Node (SBN) of NASA's Planetary Data System (PDS) \citep{McLaughlin2014-9P}.}
\addtocounter{footnote}{1}
\footnotetext[\thefootnote]{Credit: Courtesy of Silvia Protopapa through private communications. The panels represent composition maps of the coma of comet 103P/Hartley~2 published by \cite{Protopapa2014} and derived from data publicly available through the Small Bodies Node (SBN) of NASA's Planetary Data System (PDS) \citep{McLaughlin2013}.}
\addtocounter{footnote}{1}
\footnotetext[\thefootnote]{Credit: Courtesy of David Kappel through private communication. The panel represents a map derived from the current state of calibration of the data (cube I1\_00388776027) which is publicly available through the European Space Agency's Planetary Science Archive (PSA, \url{https://www.cosmos.esa.int/web/psa/rosetta}). This data was first published by \cite{Fink2016Icar}.}

\section{\textbf{Emergent coma structures}}\label{sec:emergentStructures}
Often gas structures are more difficult to observe than the dust.
Because the dust is coupled to the gas (see Eq.~\ref{eq:dust-motion-eq}) dust particles -- to some extent -- trace the gas flow.
There are meaningful differences between the two flows which we will highlight.
Nevertheless, dust observations are often used to also inform our understanding of the gas flows.

Since the first up-close observations of 1P/Halley in 1986 collimated gas and/or dust features have been detected in the coma of all comets visited by spacecraft (Fig.~\ref{fig:gasAndDustStructures}).
As described in Sec.~\ref{sec:structures} we will refer to the observed structures in the inner coma as ``filaments'' or ``collimated features'' rather than ``jets'' unless the latter is clearly warranted.

The simplest explanation for the observed inner coma structures is the following.
One might imagine that a surface is composed of active and inactive areas.
Therefore, the active areas produce high-density areas in the coma above, which are contrasted by low-density areas over inactive surface patches.
An observed difference in coma density should thus be interpreted as the result of heterogeneity of the nucleus.
Though a compelling story, it has been known to be wrong -- at least in general --  for quite some time \citep[e.g.,][]{Crifo2004}.

\begin{figure}[ht!]
\begin{center}
\includegraphics[width=0.364\textwidth]{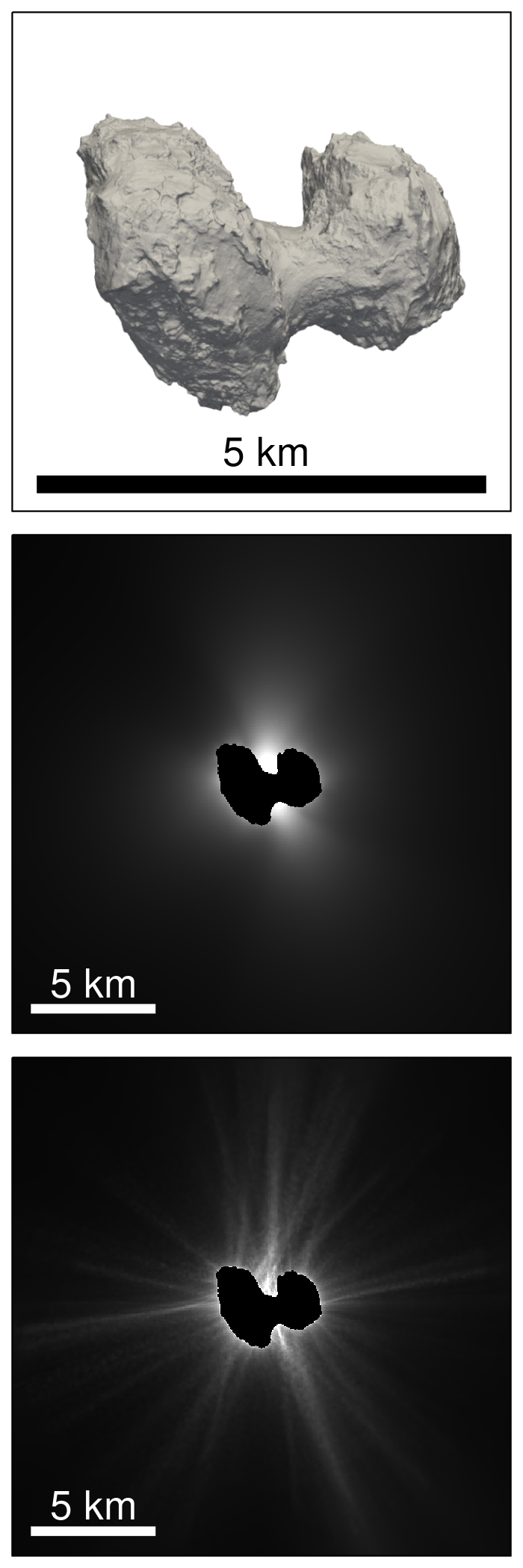}
\caption{The top panel shows the shape model \citep{Preusker2017, Jorda2016} and the orientation of comet 67P/Chryumov-Gerasimeko used in the lower two panels. The centre panel shows the water column density of 67P if it were uniformly outgassing, i.e. equal gas production rate per unit surface over the entire surface. The bottom panel shows the dust brightness of particles with a diameter of 4~$\mu$m also injected uniformly in the flow of the centre panel. Figure adapted from \cite{Marschall2017PhD}\protect\footnotemark.}
\label{fig:uniformEmissionStructures}
\end{center}
\end{figure}

There are at least two mechanisms that can produce collimated features in the absence of a heterogeneous nucleus.
One important thing to note, is that the gas structures are generally broader than the intricate dust structures (Fig.~\ref{fig:gasAndDustStructures} and \ref{fig:uniformEmissionStructures}; and \citet{Combi2012}).
Dust particles (unless they are very small [sub-micron], or in a very dense gas flow), are only weakly coupled to the gas flow.
Their primary flow direction is governed by the gas environment very close to the surface (see Eq.~\ref{eq:dust-motion-eq}) where the gas flow is dominantly perpendicular to the surface normal.
With increasing distance from the nucleus and the associated decrease in gas density, the dust particles decouple from the gas flow rather quickly.
This enables the dust to better preserve the near-surface conditions. 
For non-uniform outgassing the dust may also return and fall back to the nucleus surface \citep{Thomas2015b,Davidsson2021Icar}.
See Agarwal et al. in this volume for more detail on the dust dynamics.

In the following we will discuss two main causes leading to ``collimated features'' in the inner coma that has been identified: 1) a heterogeneous nucleus, and 2) non-sphericity of the homogeneous nucleus and local topography.
In addition to these two causes the reader should also be aware that the viewing geometry can create the illusion of structures (e.g., \citet{Shi2018}, \citet{Tenishev-2016-MNRAS} and Agarwal et al. in this volume).

The two end members of the models we will discuss are i) spherical nuclei with active areas and ii) homogeneous nuclei with complex shapes.
The former case with a spherical nucleus is agnostic as to what causes the differences between active and inactive areas.
These differences can be caused by a heterogeneous nucleus or evolutionary processes that alter the (near-)surface properties but leave the bulk nucleus properties unaltered.

\subsection{Heterogeneous nucleus surface}
A significant amount of work was performed to explore active spots on or inhomogeneous outgassing from spherical nuclei \citep[e.g][]{KomleIp1987, Kitamura1990, Knollenberg1994, Crifo1995, CrifoRodionov1997}.
In these models, a spherical nucleus would be divided into active and inactive surface elements.
Though early models \cite[e.g.,][]{Kitamura1986,Kitamura1987,Kitamura1990,KorosmezeyGombosi1990} did not use an underlying thermophysical model to determine the gas production rate at the surface, their conclusions still hold.
These studies showed the formation of the shock structures due to interactions of several ``jets''.
Even a single pure gas jet expanding into a co-current flow produces a shock structure.

Thus, as supported by our intuition, an active area will produce a gas structure in the inner coma and the simple story we discussed above is one way of explaining the inhomogeneous features.
Areas of enhanced activity have indeed been identified.
Regional variations of the surface activity have been invoked to explain the gas coma features at comets 9P \citep[e.g.,][]{Finklenburg2014}, 103P \citep[e.g.,][]{Fougere2013}, and 67P \citep[e.g.,][]{Fougere2016a,Marschall2016}.
For example, \cite{Finklenburg2014} was unable to reproduce the observed data from comet 9P/Tempel 1 with a homogeneous nucleus model.

Heterogeneous surface activity does not imply an inhomogeneous nucleus.
On the contrary, there is ample evidence for regional heterogeneity resulting from, e.g., dust re-deposition.
Both comets 103P \citep{AHearn2011} and 67P \citep{Thomas2015b} showed clear evidence of dust airfall, i.e., dust that was ejected into the coma but did not reach escape speed and thus fell back to the surface.
In the case of 103P, the dust redeposited in the neck region while 67P saw large deposits on the entire northern hemisphere in addition to the neck region called Hapi \citep{Thomas2015a}.
In both cases an increased water production from the neck region where dust was deposited points to the fact that the deposited chunks contained a significant amount of water ice but were otherwise largely depleted in the hypervolatiles \citep[e.g.,][]{Davidsson2021Icar}.
The enhanced activity of Hapi and reduced activity of other dusty deposits is also reflected in the AAF of coma models \citep[e.g.,][]{Marschall2016,Fougere2016a,Marschall2017}.
That the other northern dusty deposits are not active on the inbound leg comes from their different thermal history compared to Hapi \citep{Keller2017MNRAS}.
Hapi only reenters the Sun around aphelion and therefore retains its water ice content until the next inbound leg.
The other northern dust deposits are already illuminated on the outbound leg where they loose their water ice and are subsequently largely inactive on the following inbound leg \citep{Keller2017MNRAS}.
It cannot be ruled out though that the distinct heterogeneity between the two lobes of 103P/Hartley 2 is caused by a heterogeneous nucleus, at least there are no telltale signs for that from the gas phase.

The most probable scenario is thus the emission of water-rich dust chunks from one part of the comet (the small lobe in the case of 103P and the southern hemisphere in the case of 67P) and subsequent re-deposition on other areas of the surface.
If we begin with a homogeneous nucleus then this redistribution of material across the surface results in a natural alteration of the (near)-surface properties.
These in turn result in regional differences in the outgassing.

\addtocounter{footnote}{0}
\footnotetext[\thefootnote]{Courtesy of Raphael Marschall through private communication. The figure was adapted from data presented in his thesis \citep{Marschall2017PhD}.}

\subsection{Irregular shape and topography}
As soon as multi-dimensional models were available the effect of non-spherical shapes was examined.
In contrast to the spherical models -- where surface heterogeneities were studied -- a homogeneous nucleus was assumed.
For example a ``bean'' shaped nucleus was used to model comets 1P/Halley \citep{Crifo1997a} and comet 46P/Wirtanen \citep{Fulle1999, Crifo1999a}.
A homogeneously outgassing nucleus with a ``Haser''-type gas model of 67P was used in \cite{Kramer2016ApJL} to fit dust structures.

These early models demonstrated very clearly that the focusing of the gas and dust flow from an irregular shape of the nucleus results in a strongly inhomogeneous inner coma and ``collimated structures''.
The effect of a realistic shape for comet 1P/Halley was studied in \cite{Crifo2002bAAF}. 
In this case, the study included homogeneous and inhomogeneous (set of spots) emissions.
This study showed that a more detailed description of the shape leads to a considerably more complicated flow structure near the surface. 
Most notably, the geometrical effects of the surface can be stronger than the effect of a surface inhomogeneity.

We can further illustrate this effect using a toy model of comet 67P (Fig.~\ref{fig:uniformEmissionStructures}).
In this model, the emission of gas and dust on the entire surface is uniform (i.e. the production rate per unit area is constant across the surface).
Gas ``jets'' are visible flowing upwards and downwards from the neck (centre panel of Fig.~\ref{fig:uniformEmissionStructures}) due to the focus of the flow between the two lobes.
The fine structures in the inner dust coma (bottom panel of Fig.~\ref{fig:uniformEmissionStructures}) are even more striking.
None of the many dust filaments or gas jets has a ``source'' in any meaningful sense of the word.
The observed structures are on the scale of the nucleus size and therefore only accessible to spacecraft.

This implies that absent any modelling of the gas and dust flow we cannot and should not draw any conclusions as to the ``origin'' or ``source'' of any structure observed in the inner coma as tempting as it may seem.
Unfortunately, this also suggests that inverting or tracing back of features from the coma to the surface is bound to be futile because it makes the implicit assumption that sources exist in the first place.
Only a forward modelling approach can properly evaluate the different scenarios -- heterogeneous surface vs. topographic focusing -- and thus, e.g., exclude the possibility that features are produced by the irregular shape of the nucleus.

\subsection{Interpretation of spacecraft data}
The previous two sections have left us in a sort of limbo.
We have identified two end-member scenarios both of which are capable of explaining collimated inner coma structures.
But which scenario dominates in comets?
At the time of the Comets II book, insufficient data was available to clearly answer this question and thus \cite{Crifo2004} left it open.

In the following paragraphs, we will review what has been found for the four comets visited by spacecraft for which such an analysis has been done.
We will discuss the results in chronological order, thus starting with ESA's Giotto, Japan's Sakigake, and the Russian Vega flybys at comet 1P/Halley in 1986, followed by NASA's Deep Impact/EPOXI encounters of 9P/Tempel 1 and 103P/Hartley 2 in 2010 and 2014 respectively, and finally the escort of comet 67P/Churyumov-Gerasimeko by ESA's Rosetta mission from 2014-2016.

For details about the composition of comets we refer the reader to Biver et al. in this volume.

\subsubsection{1P/Halley}
\cite{Knollenberg1996} studied the overall Halley dust coma appearance during the 1986 Giotto flyby (Fig.~\ref{fig:gasAndDustStructures}a). 
They found a gas and dust distribution with two circular active areas which result in the two observed main jet-like features roughly directed towards the Sun.
They argue that data is well explained by three ``jets'' superimposed on a weak background.
As mentioned above this is not strictly self-consistent because the ``jet'' would interact with each other and could form interaction zones or even shock fronts depending on the level of activity.
Furthermore, this model relies on axis-symmetric solutions thus prohibiting the exploration of shape effects.

\cite{Crifo2002bAAF} improved on previous work by using the nucleus shape model derived from the Vega 1 probe \citep{Szego1995}. 
They pointed out that the observed distribution can be explained primarily due to the effects of the shape.
The orientation of the nucleus was not known well enough to fully constrain the model thus introducing some ambiguity.
Though the direction of some features was not completely matched, it did show for the first time that the nucleus shape can be sufficient to explain the coma morphology without the need for any ad hoc active spots on the surface.

This difference in interpretation highlights the difficulty in determining which of the two end members is the driving mechanism.
We will see though, that the data from 9P, 103P and 67P suggest that comets don't fall in one of the two extreme cases.
Rather they appear to be driven by both regional differences in the surface activity and significant topographic modifications of the flow fields.

Another notable observation at 1P/Halley is the rapid transition to free radial outflow, within roughly 100~km ($\sim 20$ nucleus radii), suggesting no notable extended source of the dominant water molecule or another altering process in the immediate vicinity of the comet \citep{Thomas1988}.
There is a substantial discussion on extended/distributed sources at comet Halley, especially for CO \citep{Eberhardt1986}, however, the measured distribution can also be explained by a change of the production rate \citep{Rubin2009} given that the flyby covered a large range of cometocentric distances and phase angles.
A good review can be found by \cite{Cottin2008}.
Distributed sources in minor species may hence not be at odds with H$_2$O coming mostly from the nucleus.


\subsubsection{9P/Tempel 1}
To date, comet 9P/Tempel 1 is by far the most spherical comet observed up close.
It thus provides a more favourable opportunity to disentangle shape from surface composition effects.
\cite{Finklenburg2014} found that a homogeneous surface composition was not sufficient to explain the distribution of $\mathrm{H_2O}$ vapour in the inner coma.
It over-predicted the amount of vapour in the coma.
Instead, they required active areas which were broadly in line with \cite{Farnham2013} and \cite{KossackiSzutowicz2008} but required additional nightside activity at the northern pole.
Though no direct link with morphology is made, the active areas identified by \cite{Farnham2013} correspond to steep slopes and the edges of smooth areas.
\cite{KossackiSzutowicz2008} argue for a varying dust mantle thickness.
These two interpretations are not mutually exclusive. 
Steeper slopes might simply have thinner dust covers because dust cannot accumulate on them.
In this sense, the conclusions of the three above-mentioned studies are in agreement.

\cite{Gersch2018ApJ} used an asymmetric model to improved the production rates of $\mathrm{H_2O}$ and $\mathrm{CO_2}$.
They were unable to fit the spectra assuming the coma was optically thin.
This indicates that the coma need to be treated as optically thick.
Using the optically thick assumption, they found improved production rates that were almost $50\%$ larger than those derived under the assumption of optically thin conditions.

\subsubsection{103P/Hartley 2}
Comet 103P/Hartley 2 is a particularly interesting case because of its hyperactivity.
The measurements obtained by the EPOXI mission \citep{AHearn2011} indicate that large chunks, rich in water ice, are ejected and then redeposited in the topographically low region of the neck \citep{Kelley2013}.
Such large chunks can retain much of their water ice \citep[e.g.,][]{Davidsson2021Icar} and thus serve as the source of the water vapour above the neck.
This interpretation is in line with coma models, which have found good agreement with the data assuming a regionally changing surface composition.
Additionally, \cite{Protopapa2014} observed micron-sized icy grains ejected from the small lobe.
There is a strong correlation between the water ice particles, dust particles, and the $\mathrm{CO_2}$ spatial distribution.
This suggests that $\mathrm{CO_2}$ drives both the activity of dust particles and icy grains, which subsequently sublimate in the coma \citep{Protopapa2014}.

\cite{Fougere2013} found that pure emission driven by the variation in the solar incidence angle did not fit the data.
Rather enhancements of roughly one order of magnitude over the background gas emission were needed for both $\mathrm{H_2O}$ and $\mathrm{CO_2}$.
To fit the data, the emission of $\mathrm{H_2O}$ from the neck needed to be increased.
For $\mathrm{CO_2}$ the same was true for the small lobe.
Further, from the areas where $\mathrm{CO_2}$ is emitted the sublimation temperature for $\mathrm{CO_2}$ was to be assumed as the surface temperature.
Most of the icy grains get pushed to the night side by radiation pressure and lateral gas expansion which further leads to a drop of the gas density above active spots faster than $1/r^2$.
Furthermore, nucleus gravity supports this process by pulling particles back from the sub-solar direction of emission.
Most of the water contribution to the extended source occurs towards the nightside.
This is consistent with observations by \cite{KnightSchleicher2013} who reported an enhancement in OH in the anti-sunward direction.
Similar observations were made by \cite{Combi2011ApJL} using Ly$\alpha$ emission of hydrogen by SOHO/SWAN, by \cite{Bonev2013Icar} using long-slit spectra of $\mathrm{H_2O}$ emission acquired with NIRSPEC/Keck 2, and by \cite{Meech2011ApJL}.

\begin{figure*}[ht]
\begin{center}
\includegraphics[width=\textwidth]{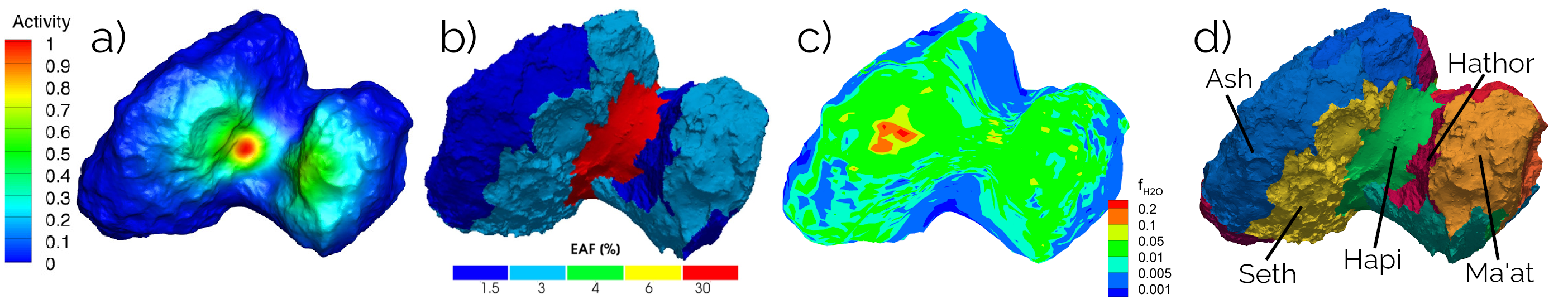}
\caption{
Panels a-c all show a variation of the active area fraction (AAF, see Sec.~\ref{sec:acitveAreaFraction}) from three different research groups modelling the same epoch of Rosetta data (Sept.-Dec. 2014). Although the scales cannot be directly compared the relative differences are. To derive production rates from these maps one needs to convolve them with the local illumination condition and respective thermal model of the surface. Panel a) shows the results by \cite{Fougere2016a}\protect\footnotemark, panel b) by \cite{Marschall2016,Marschall2019}\protect\footnotemark, and panel c) by \cite{Zakharov2018}\protect\footnotemark. Panel d) shows the morphological regions defined by \cite{Thomas2015a}.
}
\label{fig:activeAreaFraction}
\end{center}
\end{figure*}
\addtocounter{footnote}{-2}
\footnotetext[\thefootnote]{Credit: Fougere et al., A\&A, 588, A134, 2016, reproduced with permission © ESO.}
\addtocounter{footnote}{1}
\footnotetext[\thefootnote]{Reprinted from Icarus, 328, 104–126, Marschall R., Rezac L., Kappel D. et al., Copyright (2019), with permission from Elsevier.}
\addtocounter{footnote}{1}
\footnotetext[\thefootnote]{Credit: Zakharov et al., A\&A, 618, A71, 2018, © ESO, published by EDP Sciences, under the terms of the Creative Commons Attribution License (http://creativecommons.org/licenses/by/4.0).}

\subsubsection{67P/Churyumov-Gerasimeko}\label{sec:67P}
Because of the extended coverage provided by the Rosetta mission we have a very detailed picture of comet 67P/Churyumov-Gerasimeko.
The comet revealed a strong seasonal cycle caused by the obliquity of 56$^{\circ}$ \citep{Mottola2014}.
For most of the orbit, the Sun illuminates the northern hemisphere resulting in cold northern summer.
But, most of the activity occurs during southern summer when the comet is close to perihelion (southern solstice occurs only 14 days after perihelion).

Overall, the water production rate follows closely the movement of the sub-solar latitude \citep{Fougere2016b,Combi2020}. 
Furthermore, the diurnal water activity appears to also follow the sub-solar point with little thermal lag.
Dust features, driven by that activity, are sustained for tens of minutes to one hour after local sunset \citep{Shi2016}.
This also suggests that water ice, though not directly observed at the surface \citep[except for a few exposed patches;][]{Pommerol2015A&A}, resides close to the surface, likely within a few millimetres of it \citep{MarboeufSchmitt2014,Herny2021,Davidsson2021b}.

CO$_2$ and CO, in contrast to water, have much shallower diurnal cycles and more uniform emission distributions \citep{Hassig2015Sci,Combi2020}.
This behaviour is consistent with an expected longer thermal lag corresponding to deeper sublimation fronts \citep{Herny2021,Davidsson2021b}.

The water surface distribution has the strongest regional variation.
During northern summer there is a prominent dust feature originating from the northern neck region (Hapi).
Studies by \cite{Fougere2016a} and \cite{Marschall2016} both find a regional enhancement of water from that area (panels a and b in Fig.~\ref{fig:activeAreaFraction}).
\cite{Zakharov2018} on the other hand finds the enhancement at the top of the neck valley rather than deep in it (panels c in Fig.~\ref{fig:activeAreaFraction}).
But it is unclear if such a source -- located outside of Hapi -- would be consistent with the dust features or the VIRTIS gas maps \citep{Migliorini2016}.
This degeneracy illustrates that the derivation of emission maps solely from in-situ gas density data will retain a degeneracy of the solution above the previously mentioned theoretical resolution limit.
Any reconstruction method suffers from limitation, be it the iterative approach by \cite{Zakharov2018}, the spherical harmonics one by \cite{Fougere2016a} and \cite{Combi2020}, or the one using morphological regions as "basis" vectors by \cite{Marschall2016}.
Therefore, even with state-of-the-art forward models a multi-instrument approach with complementary data (e.g., in-situ gas density from an instrument like Rosetta/ROSINA in combination with column density maps from an instrument like Rosetta/VIRTIS or Deep Impact/HRI-IR) needs to be taken to break this ambiguity \citep[e.g.,][]{Fougere2016a,Marschall2019}.
Figure~\ref{fig:activeAreaFraction} also illustrates the correlation between activity and morphology.
In particular, \cite{Fougere2016a} and \cite{Marschall2019} place the maximum of activity in the Hapi region while the other regions which are dominated by dust deposits (such as Ash, Seth, and Ma'at) are at least one order of magnitude less active.

\cite{Vincent2016} hypothesized that the water activity and by extension the dust features originate primarily from cliffs.
This seems to be confirmed by coma models \citep{Marschall2017}.
Interestingly, the dusty deposits outside the Hapi region do not seem to contribute to water activity \citep{Marschall2017}.
This implies that the airfall material in those regions mostly loses its volatile content after deposition on the outbound leg of the comet's orbit after perihelion.
The deposits in the Hapi region on the other hand have preserved their water ice because they remain continuously shadowed.
These differences in activity are thus not due to inherent differences in the composition of the nucleus but stem from material transport across the surface combined with a different thermal history which alters the near-surface composition of the deposits.

Similar to the pattern observed at 103P, activity shapes the surface composition and thus in turn the observed activity.
There are regions on comets that do not simply erode continuously and thus are not purely driven by the nucleus composition but rather by the feedback of redeposited material.

Most of the dust features overlaying this regional variation of activity appear purely driven by local topography \citep[e.g.,][]{Marschall2016,Marschall2017} or the viewing geometry \citep[e.g.,][]{Shi2018}.


\subsubsection{What causes the observed coma structures?}
We thus return to the question of what causes the observed coma structures.
There is clear evidence from all visited comets that there are regional differences in the strength of activity.
These regional differences though are consistent with evolutionary processes and material transport across the surface.

Large dust particles or chunks that do not reach escape speed will redeposit as airfall on the surface filling regions of gravitational lows.
The angle of repose seems to be at roughly $30^{\circ}$ \citep{Vincent2016,Marschall2017} corresponding to granular material.
This leaves cliffs free of such insulating airfall and thus active \citep[e.g.,][]{Farnham2013,Vincent2016,Marschall2017}.

The local topography and overall shape of the comet significantly shape the flow \citep[e.g.,][]{Crifo2002bAAF,Shi2018,Marschall2016}.
The effect of topography is even more pronounced in the dust which does not seem to require any additional small-scale (i.e. below the resolution limit of the gas) localized sources.
A series of work \citep[e.g.,]{Crifo2003, Zakharov2008, Zakharov2009} even showed that flow structures generated due to surface topography and inhomogeneity can be identical, i.e. making it impossible to derive the reason for the structure formation just from how they appear.
Therefore, it is crucial that additional information, like surface morphology and composition, is taken into account when assessing the cause of the structures in the coma.
Features arising from a ``flat'' topography would indicate inhomogeneous sources.
In contrast, a source region which is rather uniform in morphology (or, e.g., spectral properties) but has significant topography points to the topography as the source of those features.

Currently, most results point to the fact that no heterogeneous nucleus is needed to explain the heterogenous coma.
Further, apart from outbursts, there is no evidence for jets in the traditional sense of having a confined source area.
Rather, there appears to be smooth activity originating from large regions.
The absence of any observable shock features also points in this direction.

We should remind the reader of the fact that determining the surface-emission distribution from in-situ gas measurements that are tens to hundreds of nuclei radii removed from the surface has a physical resolution limit of several hundreds of meters \citep{Marschall2020a}.
This can hide smaller scale heterogeneities in the activity maps derived from such measurements.

From an Occam's razor argument we can be satisfied with the above conclusions but future measurements should never the less probe the issue of sources below the physical resolution limit mentioned above.
We will discuss which measurements can address this issue in Section~\ref{sec:outlook}.

It appears -- at least at this point -- that there is no need for more refined activity maps to explain the coma data.
Regional differences of the strength of outgassing (i.e., AAF) in combination with local changes in illumination and topography are sufficient to predict the coma structures.

\begin{figure*}[ht!]
\begin{center}
\includegraphics[width=0.95\textwidth]{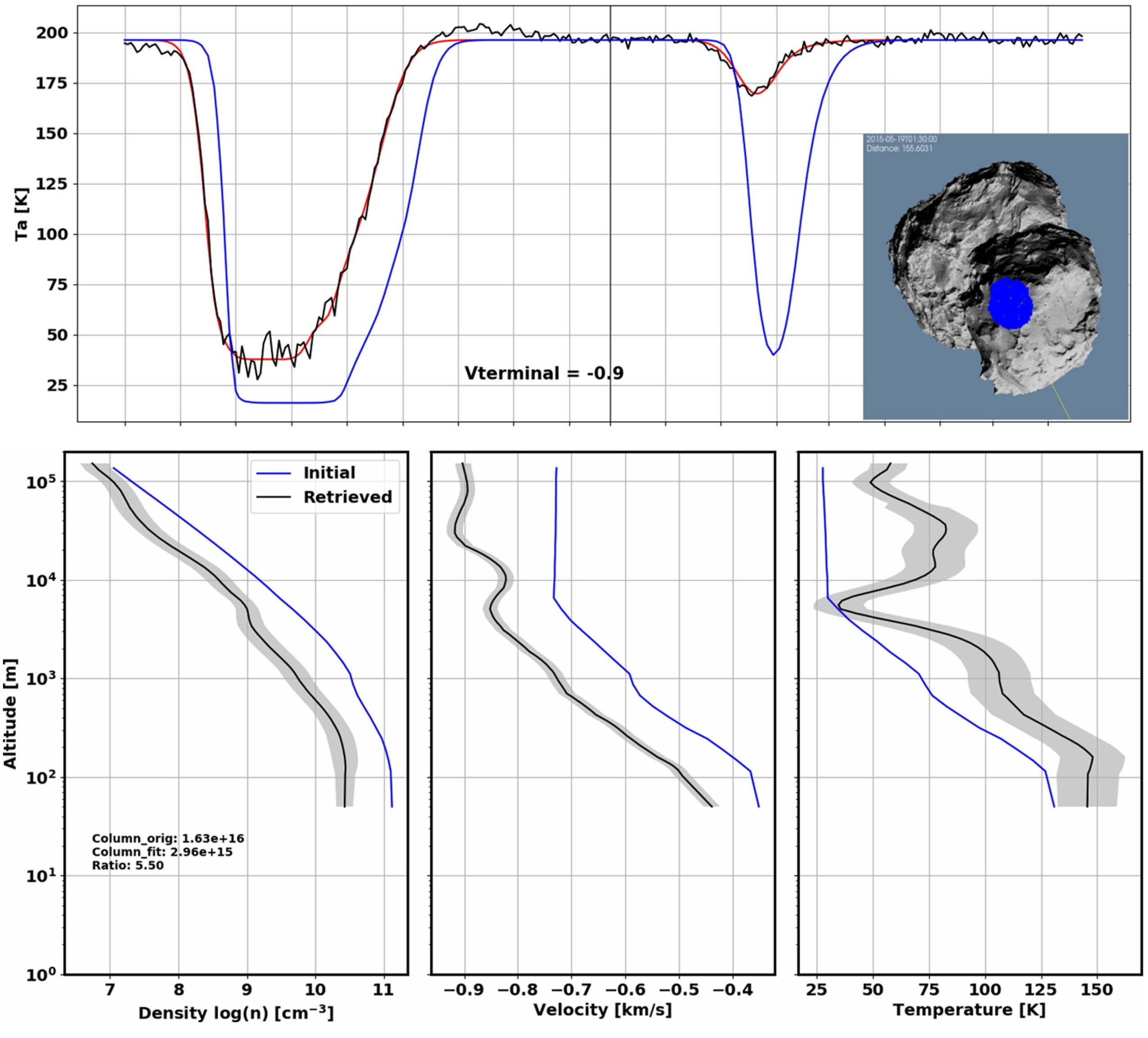}
\caption{
The top panel shows three spectra for the 557 GHz transitions of H$_2^{16}$O (left) and H$_2^{18}$O (right) for an observation on 2015-05-19T01:30: (black) MIRO measurement, (red) retrieved synthetic spectrum and (blue) synthetic spectrum of the gas model. The inline-image shows the shape model of 67P and the pointing geometry with the MIRO footprint (blue circle) on the nucleus surface. The bottom three panels show the vertical profiles of the number density (left), the expansion velocity along the MIRO line-of-sight (middle), the kinetic temperature (right) for the gas model in blue, and the retrieved profiles in black lines. The shaded region represents a 2$\sigma$ component of uncertainty due to measurement random error propagation.
The figure is adapted from \cite{Marschall2019}\protect\footnotemark.
}
\label{fig:miro}
\end{center}
\end{figure*}
\addtocounter{footnote}{0}
\footnotetext[\thefootnote]{Reprinted from Icarus, 328, 104–126, Marschall R., Rezac L., Kappel D. et al., Copyright (2019), with permission from Elsevier.}

\subsubsection{Gas temperature and speed observations}

As discussed above, most models have used either local number densities or integrated column densities to constrain surface parameters.
The other flow properties -- temperature and flow speed -- have somewhat been neglected for inner coma models.
The main reason is that there are limited spacecraft observational constraints with one notable exception.

MIRO has provided information in addition to the number or column density of the gas flow.
Due to its ability to detect the H$_2^{16}$O and H$_2^{18}$O absorption lines information about the water speed and temperature can be retrieved \citep{Biver2019A&A,Rezac2021A&A}.
\cite{Biver2019A&A} observed general agreement between the terminal velocity and theoretical expectations from \cite{Hansen2016MNRAS} over a large range of heliocentric distances (from 3.8~au to perihelion at 1.24~au).
The terminal gas speeds increased from $\sim 600$~m/s at 3.6~au inbound to $\sim 900$~m/s at perihelion and then subsequently decreased to $\sim 400$~m/s at 3.8~au outbound.
Notably, the lower terminal gas speed at similar heliocentric distances might be caused by the higher fraction of CO$_2$ in the coma outbound vs. inbound \citep{Biver2019A&A,Combi2020}.

Using VIRTIS data, \cite{Cheng2022A&A} found rotational temperatures for water of $\sim150$~K close to the surface.
At distances of 10~km from the centre of the comet the temperatures had dropped to $60-80$~K.
These observations were taken at heliocentric distances between 1.4 and 1.8~au.

Self-consistent retrievals of the line-of-sight (LOS) profiles of water number density, temperature, and velocity from the spectral lines are also possible.
For data from around the inbound equinox (May 2015), \cite{Marschall2019} found a notable deviation from the expected model profiles compared to the retrieved profiles from the data.
Although several observations were found where the gas model and the LOS retrievals were in good agreement there were others (example shown in Fig.~\ref{fig:miro}).
In some of these cases, the gas velocity along the LOS was up to 100~m/s faster in the data than in the model (bottom middle panel of Fig.~\ref{fig:miro}).
Conversely, the retrieved gas temperature from the spectrum was warmer than in the model.
This is intriguing because the model assumed that the surface gas temperature is at the free sublimation temperature ($\sim200$~K) and therefore \cite{Marschall2019} speculated that in reality, the gas must have been significantly warmer upon emission than assumed.
This could therefore imply that the gas sublimates from the sub-surface and thus first travels through a much hotter desiccated surface layer.
That gas can be efficiently heated when flowing through a porous layer has been shown for different porous surfaces \citep[e.g.,][]{Skorov2011, Christou2018, Christou2020P&SS}.

What this demonstrates is the potential information contained in the gas speed and temperature in the near nucleus environment.
These kinds of measurements contained, e.g., in the Rosetta/MIRO dataset remain largely unexploited.
Recent work using a thermophysical model, \cite{Davidsson2022MNRAS} demonstrates that $\mathrm{CO_2}$ ice is fairly close to the surface for its effects to be detected by MIRO \citep{Davidsson2022MNRASb}.

The determination of the gas temperature can also be done from ground.
For comet 73P-B/Schwassmann–Wachmann 3 \cite{Boncho2008Icar} detected multiple $\mathrm{H_2O}$ emission lines in non-resonant fluorescence near $2.9~\mu$m using the Subaru telescope.
They retrieved a decrease in the rotational temperature from $\sim110$ to $\sim90$~K as the projected distance from the nucleus increased from $\sim5$ to $\sim30$~km.
These measurements were taken when the comet was at a heliocentric distance of $1.027$~au.
\cite{Fougere2012} used this data to compare it to their kinetic coma model.
Importantly, \cite{Fougere2012} find that the comparison with a model using pure water emission from the surface cannot account for the observed rotational temperatures.
Such a model would predict a much steeper drop of the water column density and temperature with distance to the nucleus than was observed for comet 73P.
By introducing icy particles into the model -- acting as an extended gas source -- both the temperature and column density of the gas could be increased at large cometocentric distances to match the observations.
With the extended source model \cite{Fougere2012} was able to conclude that the water coma of 73P is dominated by sublimation from icy grains in the coma rather than surface sublimation.

\section{\textbf{Outlook and open questions}}\label{sec:outlook}
As described above, in-situ measurements taken beyond several nucleus radii from the surface are limited by the physical resolution limit stemming from the gas dynamics itself \citep{Marschall2020a}.
This is made even worse by data primarily from terminator orbits, as often the case during the Rosetta mission, because of the lateral day-to-nightside flow of the gas in this region.
This naturally leads to the question: Which observations can constrain the gas source distribution at the surface better?

First, in-situ coma measurements at small phase angles, where the gas flow is mostly radial after only a few kilometres even for an irregularly shaped nucleus, would provide stronger constraints on the surface source distribution.
Second, in-situ coma measurements at much lower altitudes (a few hundred meters) above the surface would break the degeneracy stemming from the resolution limit.
Such measurements have the added benefit of being within the acceleration region of the gas thus providing valuable constraints on the velocity distribution function at the surface (if the gas temperature and speed can be measured in addition to the gas density) and any additional processes within that region, such as the sublimation of small icy particles, or significant mass loading.
Third, high spatial resolution ($<10$~m) spectral imaging could probe the near-surface structure of the gas flow without the need of flying a spacecraft close to the surface.
In this case, a terminator orbit would be favoured to show the emission into the sub-solar direction.
Fourth, measurements at a high phase angle would allow the determination of the amount of night side activity, which is currently rather poorly constrained.
Such data could also shed more light on the fading of activity after local sunset and thus hold valuable information about the depth of sublimation fronts and thermal properties (e.g., thermal inertia) of the subsurface. 
\\

The next open question pertains to the surface boundary conditions of coma models, i.e. the nucleus surface.
What is the gas temperature and velocity distribution function at the nucleus surface?
While the gas number density at tens of nucleus radii is rather insensitive to the surface gas temperature and velocity \citep[e.g.,][]{Liao2016} the temperatures and velocity themselves are not.
There is some hope, that existing data (e.g., from Rosetta/MIRO), can still provide important information \citep[e.g.,][]{Marschall2019,Pinzon2021} about the near nucleus structure of the flow and by extension of the surface.
But such advances will require the implementation of an actual non-idealized thermal model of the subsurface to inform the gas coma model boundary conditions.
A real thermal model will be crucial to properly remove diurnal/seasonal variability of the production rates and retrieve (sub-)surface properties.
\\

While the above-mentioned measurements illustrate a way forward from the standpoint of in-situ and remote sensing data of the coma, they also highlight the greater problem that is only addressed indirectly by these measurements.
It is the big overarching question: What drives and sustains cometary activity?
It has become clear, that even the Rosetta mission, with its extensive data set, was not able to answer this question \citep{KellerKuehrt2020}.
And it can be doubted whether further in-situ and remote sensing data from the coma, even with better instruments and/or different observation geometries, will be able to adequately answer this question.
The lesson for this ``failure'' has to be that we need to understand the physico-chemical structure of the sub-surface.
This has already been pointed out by \cite{Thomas2019SSRv} but is worth reiterating here.
The straightforward way of understanding how activity works is by observing it in situ at multiple locations on the surface and determining how volatiles and refractories are mixed on the microscopic level.
We thus deem it inevitable that a comet lander or hopper is needed to comprehensively address this question.
We should also mention that it is unlikely that any sample return mission would be able to answer this question. 
Any sample returned to Earth would not be able to retain the physical structure of the sample during the high accelerations during re-entry.
Furthermore, retaining the ices of highly volatile species is very challenging. 
Sublimation of such molecules may further alter the physical structure.
And it is this physical structure and how volatiles are embedded within the refractory components that seem to hold the key to understanding cometary activity.
\\

A final open question we want to highlight touches on the outer coma and tail.
Though these were not the focus of this chapter it is worth pondering them.
Now that we know there are regional differences in surface activity, modelled without the need for any small-scale localised sources (in the sense of a classical ``jet'') it is worth asking what the consequences are for the structures we see in the outer coma.
First, it is more obvious than ever that the concept of a ``jet'' in the outer coma is a complete misnomer \citep{Crifo2004}.
If there are only large regional differences in the level of activity but no source in the traditional sense, then such features in the outer coma structures cannot be ``jets''. 
Second, the apparent mismatch between the intricate and rather small-scale structures and the comparably large spatial extents of outer coma ``jets'' needs to be resolved.
How do these small-scale structures of the inner coma connect to the outer coma?
We have yet to understand how coma structures connect from the surface to the inner coma (accessible to spacecraft) and ultimately to structures in the outer coma (accessible with ground-based telescopes).


\section{\textbf{Conclusions}}\label{sec:conclusions}
In this chapter we have focused on two crucial questions of linking inner coma measurements to the surface of cometary nuclei:
\begin{enumerate}
    \item How can we derive the gas production rate of different species and thus the volatile mass loss from coma measurements?
    \vspace{-0.2cm}
    \item Can we determine if coma structures (inhomogeneities in density, often referred to as ``jets'') are reflective of a heterogeneous nucleus, or are mere emergent phenomena in the gas flow due to, e.g., the complex shape of the nucleus?
\end{enumerate}

Answering these questions allows us to i) link spacecraft measurements from the inner coma to the surface which is also a prerequisite to understanding ground-based observations and linking those measurements to the nucleus, and ii) make predictions for future comet missions and assess hazards for spacecraft operating in that region.

Deriving properties of a cometary nucleus from coma data is of significant importance for our understanding of cometary activity and has implications beyond.
For example, whether a cometary nucleus is homogeneous or heterogeneous in composition will influence how we understand planetesimal formation.
A homogeneous nucleus would indicate that the material was formed in close proximity and within a rather short time.
A mechanism such as the streaming instability \citep[e.g.,][]{Goldreich1973ApJ, Youdin2005ApJ, Johansen2007Natur, Simon2016ApJ, Blum2017MNRAS} followed by gravitational collapse would likely result in rather homogeneous nuclei.
In contrast, a heterogeneous nucleus would support longer formation times which allow for more mixing and therefore might be indicative of hierarchical formation scenarios \citep[e.g.,][]{Weidenschilling1997Icar, Kenyon1998AJ, Windmark2012aA&A, Windmark2012bA&A, Davidsson2016A&A}.

To understand the coma-surface link, different types of models have been developed over the past decades. 
These have been used to derive properties of the nucleus surface from in-situ or remote sensing instruments of spacecraft.
We find those common heuristic models such as the ``Haser model'' (and more complex variations thereof) are useful to derive properties such as the global gas production rate.
We caution the reader though that they are not a substitute for real physically self-consistent models. 
Heuristic models are -- by their nature of neglecting physical processes -- limited and therefore not well suited to, e.g., derive the gas emission distribution from the surface.
Such models cannot give us answers as to the emergence of structures in the coma and properties of the (sub-)surface.

To retrieve more detailed information about the dynamical structure of the gas coma current state-of-the-art physical models are needed.
We described the need for accurate thermophysical models of the (sub-)surface that serve as input for state-of-the-art kinetic coma models.
Typically, kinetic models developed within the DSMC method are used to model the dusty gas dynamics in cometary comae.
Although they are computationally more expensive they are our best tool to derive the near-surface properties.
This includes, e.g., the determination of the gas temperature and speed when it leaves the surface.
The gas temperature and speed hold information about the physio-chemical sub-surface structure that the gas flows through before leaving the nucleus.
Therefore, we encourage the reader to opt for forward models whenever possible, because they are less likely to be misinterpreted and their limitations forgotten.

Complex models, which account for all relevant processes, are particularly useful when there are non-linear effects (such as molecular collisions) which cannot easily be parameterized.
If a flow contain only linear features, or if observations only capture global properties of the flow, then complex models do not offer a benefit over simplified models.
This is the case, for instance, when determining the global gas production rate.
At the same time, we should always keep in mind that all models have some limitations.
A complex model, such as DSMC, is more widely applicable but usually comes at a computational cost.
Thus, understanding these limitations prevents us from misapplying the model to inappropriate situations (see, e.g., discussion in Sec.~\ref{sec:regimeEstimates}   and \ref{sec:kineticDynamicModels} with respect to the flow regime).

One key lesson to be learned from physical models is that there is a physical resolution limit to determining the distribution of source regions on the nucleus surface, especially from in-situ data taken at tens to hundreds of kilometres away from the surface.
Sources at the surface do not independently/linearly contribute to the gas density in the coma.
This limits the accuracy to which we can resolve the heterogeneity at the surface, both with physical and heuristic models, even if the latter are mathematically valid to smaller spatial scales.

One of the main open questions from the previous volume \citep{Crifo2004} was what drove ``jet'' like structures in the coma.
Are they the result of i) a heterogeneous nucleus or ii) features emerging from the focusing of flows from a complex nucleus shape and local topography?
The past decades of spacecraft missions to comets 1P/Halley, 19P/Borrelly, 9P/Tempel 1, 67P/Churyumov-Gerasimeko, 81P/Wild 2, and 103P/Hartley 2 have given important insights to this question.
The nuclei show a vast diversity in shape with both concave and convex terrain.
It has become clear that all the nuclei visited show regional differences in the strength of outgassing beyond simple variations in local illumination.
These regional differences though do not appear to be driven by a heterogeneous nucleus itself but rather by surface processes that alter the near-surface composition.
One of the main drivers of this surface evolution is airfall -- the re-depositing of large dust chunks that are depleted in ices more volatile than water. 
In addition to the regional variations of activity, local topography and the global complex shape of the nuclei is sufficient to explain the observed inhomogeneous structures of the inner coma.
In most examples, there does not appear to be any need for small-scale sources that would link to ``jets''.
This excludes, of course, the phenomena of outbursts that have a clear confined source location at the surface.

Thus while there is no strong evidence for heterogeneous nuclei, heterogeneous nucleus surfaces are a common feature of comets.
Real comets thus don't neatly fall into one of the two extremes outlined above.
Rather they are an intermediate case, i.e., regionally heterogeneous surfaces with corresponding topography shaping an inhomogeneous inner coma.

This also illustrates further, what has been known for some time, that the term ``jet'', which has a strict physical meaning, is a misnomer \citep{Crifo2004,Vincent2019}.
Because the ``jet'' is not an adequate description of the emergent structures in the inner coma, it also loses its meaning in the outer coma.
Thus there is a need for a consistent, widely accepted nomenclature that more accurately describes the structures and features in cometary comae, near and far.

To improve the resolution of surface emission maps we propose that orbits, much closer to the surface (several hundred meters), and at low ($<30^{\circ}$) and high ($>150^{\circ}$) phase angles are needed to derive more accurate surface properties from in-situ coma measurements.
Such a data set would also allow us to probe the acceleration region of the gas, and determine the level and nature of the night-side activity. 
Alternatively, high spatial resolution spectral imagers that can probe the coma structure very close to the surface from a terminator orbit, would also significantly improve our understanding of the link between coma structures and the surface while remaining further away from the surface.
Measuring the gas temperature and speed close to the surface is also crucial to deriving (sub-)surface properties.
This can be done with an instrument like Rosetta/MIRO, or in-situ on low-altitude orbits.

We are now at a point where we start to understand the near nucleus environment well enough that a crucial missing element can be tackled.
How do coma structures on a spatial but also temporal scale trace from the surface out to the scales observable with ground-based telescopes?
Answering this question will allow us to understand the nature of the large-scale structures in the outer coma, their origin, and what they tell us about the nucleus.

Finally, we suggest that observations of activity at the surface in addition to sub-surface measurements are likely the key avenue to definitively address how comets work by probing the physio-chemical structure of the sub-surface.
In addition, e.g., thermal measurements, radar sounding, and subsurface sampling and imaging (to determine the physico-chemical properties), will all play an important role to understanding comets \citep{Thomas2019SSRv}.
These goals can be achieved by dedicated orbiter and lander infrastructures.
While gas coma measurements close to the surface can shed important insights into the sub-surface structure, degeneracies likely will remain.
A lander or hopper that can study the near-surface physio-chemical properties is thus the most promising next step to address this issue.


\vskip .5in
\noindent \textbf{Acknowledgments.} \\
We thank the two reviewers, Lori Feaga and Vladimir Zakharov, for taking the time to proving detailed constructive suggestions that have greatly improved this chapter.
We thank Dominique Bockelée-Morvan for taking time, as editor, to read the chapter and give us important feedback and suggestions.\\
Furthermore, we thank Lori Feaga, Silvia Protopapa, Alessandra Migliorini, and David Kappel for their help with the water and $\mathrm{CO_2}$ column density maps for 9P, 103P and 67P respectively, and providing high resolution versions of the respective figures.\\
RM acknowledges the support from NASA's Emerging Worlds program, grant NNX17AE83G, and funding from the European Research Council (ERC) under the European Union’s Horizon 2020 research and innovation programme (Grant agreement No. 101019380). \\
A portion of this research was carried out at the Jet Propulsion Laboratory, California Institute of Technology (internal “raise the bar” funding is acknowledged), under a contract with the National Aeronautics and Space Administration.

\bibliographystyle{sss-three.bst}
\bibliography{Chapter}

\end{document}